\documentclass[a4paper,11pt]{article}
\pdfoutput=1
\usepackage{jcappub,macros,graphicx,float,mathtools,caption,subcaption} \graphicspath{{./Figures/}}
\usepackage[T1]{fontenc}
\usepackage[breakable]{tcolorbox}
\usepackage{soul}
\usepackage{bigints}

\title{\boldmath Modified propagation of gravitational waves from the early radiation era}

\author[a,b]{Yutong He,}
\author[c,d,e]{Alberto Roper Pol}
\author[a,b,e,f]{and Axel Brandenburg}

\affiliation[a]{Nordita, KTH Royal Institute of Technology and Stockholm University, Hannes Alfv\'ens v\"ag 12, SE-10691 Stockholm, Sweden}
\affiliation[b]{The Oskar Klein Centre, Department of Astronomy, Stockholm University, AlbaNova, SE-10691 Stockholm, Sweden}
\affiliation[c]{D\'epartement de Physique Th\'eorique, Universit\'e de Gen\`eve, CH-1211 Gen\`eve, Switzerland}
\affiliation[d]{Universit\'e Paris Cit\'e, CNRS, Astroparticule et Cosmologie, Paris, F-75013, France}
\affiliation[e]{School of Natural Sciences and Medicine, Ilia State University, 3-5 Cholokashvili Ave, Tbilisi, GE-0194, Georgia}
\affiliation[f]{McWilliams Center for Cosmology and Department of Physics, Carnegie Mellon University, 5000 Forbes Ave, Pittsburgh, PA 15213, USA
\\ \today}

\emailAdd{yutong.he@su.se}
\emailAdd{alberto.roperpol@unige.ch}
\emailAdd{brandenb@nordita.org}

\abstract{We study the propagation of cosmological gravitational wave (GW)
backgrounds from the early radiation era until the present day in
modified theories of gravity.
Comparing to general relativity (GR),
we study the effects that modified gravity parameters, such as
the GW friction $\alpM$
and the tensor speed excess $\alpT$,
have on the present-day GW spectrum.
We use both the WKB estimate, which provides
an analytical description but fails at superhorizon scales, and numerical
simulations that allow us to go beyond the WKB approximation.
We show that a constant $\alpT$ makes relatively insignificant changes to the GR solution,
especially taking into account the constraints on its value from GW 
observations by the LIGO--Virgo collaboration, while
$\alpM$ can introduce modifications to the spectral slopes of the GW
energy spectrum in the low-frequency regime
depending on the considered time evolution of $\alpM$.
The latter effect is additional to the damping or growth
occurring equally at all scales that can be predicted by the WKB approximation.
In light of the recent observations by pulsar timing array (PTA)
collaborations, and the potential observations by future detectors such as SKA, LISA, DECIGO, BBO, or ET,
we show that, in most of the cases, constraints cannot be placed on the effects of $\alpM$ 
and the initial GW energy density $\EEGW^*$ separately, but only on the combined 
effects of the two,
unless the signal is observed at different frequency ranges.
In particular, we provide some constraints on the combined effects from the reported PTA observations.
}

\begin{document}
\maketitle
\flushbottom

\section{Introduction}
\label{sec:introduction}

The present day Hubble constant, $H_0$, is measured to be around $H_0\sim74\,{\rm km/s/Mpc}$ by astrophysical tests using type Ia supernovae~\cite{Riess:2019cxk,Riess:2020fzl,Riess:2021jrx}, lensed quasars~\cite{Wong:2019kwg}, and megamaser-hosting galaxies~\cite{Pesce:2020xfe}.
On the other hand, cosmological tests of $H_0$ constrain its value to $H_0\sim67\,{\rm km/s/Mpc}$ from
cosmic microwave background (CMB)~\cite{Planck:2018vyg} and baryon acoustic oscillations (BAO)~\cite{BOSS:2016wmc} experiments, assuming the standard $\Lam$CDM ($\Lam$ cold dark matter) model of cosmology.
These measurements suggest the presence of a $4$--$5\,\sig$ deviation
discrepancy between the early- and late-universe measurements of
$H_0$~\cite{Verde:2019ivm}, known as the Hubble tension.
The observation of gravitational waves (GWs) from ``standard sirens''
provides an independent measurement of the Hubble rate~\cite{Schutz:1986gp,Holz:2005df}.
Current observations from LIGO--Virgo present large uncertainties and cannot yet solve the tension \cite{Nissanke:2013fka,LIGOScientific:2017adf,DES:2020nay,Palmese:2021mjm}.
However, the
measurement of $H_0$ is expected to drastically improve with the
space-based GW detector LISA \cite{LISACosmologyWorkingGroup:2022jok} or the next-generation
ground-based detectors like the Einstein Telescope 
(ET) \cite{Branchesi:2023mws}.
Besides the well-known $H_0$ tension, a number of other observational discrepancies within the $\Lam$CDM model have been reported,
such as the $S_8$ tension, where $S_8\propto\sig_8\,\Om_{\mat,0}^{1/2}$ characterizes the structure growth rate $\sig_8$ and today's matter density $\Om_{\mat,0}$~\cite{Joudaki+17}.
A recent summary of over a dozen cosmological tensions and their varied
significance can be found in ref.~\cite{Perivolaropoulos+21}.

In view of these tensions, $\Lam$CDM, although being an extremely successful model, does not provide a
complete picture and it might be necessary to go beyond it.
Since general relativity (GR) is responsible for the gravity sector of $\Lam$CDM,
modifying $\Lam$CDM often requires modifying GR.

In search for modified gravity (MG) theories, 
certain systematic approaches have been developed to go beyond GR,
such as relaxing one or some of the assumptions made in Lovelock's theorem that proves the uniqueness of GR~\cite{Lovelock71,Lovelock72}.
Overarching families of MG theories have been constructed to represent a vast number of individual models,
for example Horndeski's theory, which 
corresponds to all four-dimensional scalar-tensor models with second-order
derivatives~\cite{Horndeski74,Deffayet:2011gz,Kobayashi:2011nu},
beyond Horndeski \cite{Zumalacarregui:2013pma,Gleyzes:2014dya,Gleyzes:2014qga}, Degenerate Higher-Order Scalar-Tensor (DHOST) \cite{Langlois:2015cwa,BenAchour:2016fzp,Langlois:2017mxy},
Ho\v{r}ava-Lifshitz \cite{Horava:2009uw,Blas:2010hb},
Generalized Einstein-Aether gravity \cite{Jacobson:2000xp,Zlosnik:2006zu}, standard \cite{Tasinato:2014eka,Heisenberg:2014rta} and beyond \cite{Heisenberg:2016eld,Allys:2015sht,BeltranJimenez:2016rff} generalized Proca, bimetric gravity \cite{deRham:2010ik,deRham:2010kj,Hassan:2011zd}, and tensor-scalar-vector (TeVeS) \cite{Sagi:2010ei} theories.
Some specific examples of MG theories are: Brans-Dicke~\cite{BD61}, quintessence~\cite{Ratra:1987rm}, $f(R)$~\cite{DeFelice+10}, $f(G)$~\cite{Carroll:2004de}, $k$-essence~\cite{Armendariz-Picon+00}, kinetic gravity 
braiding~\cite{Deffayet:2010qz}, and galileon~\cite{Nicolis:2008in}.
For a comprehensive review see, e.g., 
ref.~\cite{Clifton+11}.

At the level of tensor linear perturbations, MG theories introduce
three parameters to the standard gravitational wave (GW) equation that can be conveniently used as a phenomenological probe in the parameter space of MG:
\begin{itemize}
\item GW friction $\alpM$, usually\footnote{This is not necessarily the only interpretation of the GW friction; see, e.g., ref.~\cite{Belgacem:2018lbp}.} associated to the running of the effective Planck mass
$\alpM = \dd \ln M_\eff^2/\dd \ln a$.
The friction term yields a GW luminosity 
distance different than the electromagnetic (EM) counterpart $\dGW\neq\dEM$ 
that allows us to put constraints on $|\alpM|\sim\ooo(10^0)$
~\cite{Deffayet:2007kf,Saltas:2014dha,Lombriser+15,Nishizawa17,Belgacem:2017ihm,Belgacem:2018lbp}.
Signals propagating over longer distances could provide tighter constraints.
Such ``standard siren'' measurements have so far yielded mild constraints on $\alpM$,
as LIGO--Virgo covers a relatively small distance of $z\sim\ooo(10^{-1})$~\cite{Mastrogiovanni:2020gua,Mastrogiovanni:2020mvm}.
Cosmological tests involving the CMB provide a tighter constraint of $|\alpM|\lesssim\ooo(10^{-1})$~\cite{Planck:2018vyg} at the present day;
see \Sec{sec:param} for details.
This aspect can also be improved with LISA \cite{Arai+17,LISACosmologyWorkingGroup:2019mwx,LISACosmologyWorkingGroup:2022jok} and ET \cite{Belgacem:2017ihm,Belgacem:2018lbp,DAgostino:2019hvh,Matos:2021qne,Matos:2022uew,Branchesi:2023mws}, which might probe distances up to
$z\sim\ooo(10^1)$.
\item Tensor speed excess $\alpT$, related to the GW speed $\cT$ via $\cT^2 = \alpT + 1$.
It has been tightly bounded to $\alpT\lesssim\ooo(10^{-15})$ 
by the multi-messenger detection of GW170817 and GRB 170817A~\cite{LIGOScientific:2017zic}.
This observation in turn constrains the theory space of MG~\cite{Baker:2017hug,Ezquiaga:2017ekz,Creminelli:2017sry,Sakstein:2017xjx}.
However, this constraint only applies to the frequency range probed by LIGO--Virgo and might not necessarily apply at different frequencies, as proposed in ref.~\cite{deRham:2018red}.
\item Graviton mass $\mg$, which modifies the otherwise linear dispersion relation in GR.
\end{itemize}
In this paper, we focus on the study of the modifications on the 
expected relic cosmological GW spectrum under MG for different 
time-dependent GW friction models, and for constant values of $\alpT \neq 0$.
We present the general theoretical framework for time-dependent $\alpM$ and $\alpT$, but defer the
study of time and/or frequency dependent $\alpT$ to future work.
For a similar study concerning the effects of $\mg$, see ref.~\cite{He:2021bqm}.
Our focus on the cosmological effects of $\alpM$ is motivated by its less stringent constraints compared to $\alpT$,
and its accumulated effects over cosmic history.
We ignore the effects of scalar perturbations in MG that can lead to gravitational slip and focus only on the phenomenological effects
on the tensor sector.
For the combined study of scalar and tensor perturbations in MG see refs.~\cite{Saltas:2014dha,Matos:2022uew}.
Similarly, we ignore modifications on the dark energy equation of state
and set $w_{\rm DE} = -1$ (i.e., $w_0 = -1$ and $w_a = 0$) \cite{Chevallier:2000qy,Planck:2018vyg}, such that the $\Lam$CDM background expansion is unmodified.

To this end,
we initialize a GW spectrum based on previous studies of GWs sourced by primordial magnetohydrodynamic (MHD) turbulent
fields in the early radiation era---%
e.g., at the electroweak or QCD phase transitions (EWPT or QCDPT)---%
and then propagate it through cosmic history to obtain the present-day relic spectrum.
Actually, we take a generic double broken power law GW spectrum
and we show that the obtained
results are independent of the initial spectral shape.
Hence, our results can be applied to predict the expected
GW spectrum from a generic source in MG.
We then analytically and numerically compare the relic spectra in GR and MG, and discuss their potential
observational implications.
The numerical solutions in this study are obtained using the {\sc Pencil Code}~\cite{pencil},
which has been used as a tool for simulations of GWs from primordial turbulent sources since the implementation of a GW solver~\cite{RoperPol:2018sap}.
The cosmological history of the universe is modelled by numerically
solving Friedmann equations.
We provide a cosmological solver and a tutorial within the associated online material \cite{GH}.
It also contains the routines that have been used to process the data
obtained from the {\sc Pencil Code} and to generate the results and plots
of the present work.

We introduce the propagation of tensor-mode perturbations described by the 
GW equation in MG in \Sec{mod_GWs}.
In \Sec{WKB_s}, we introduce the WKB approximation commonly used to describe approximate solutions
to the modified GW equation.
We derive the WKB solution in \Sec{sec:WKB}, study its limitations and range of validity
in \Sec{limitations_WKB}, and compute the GW spectrum in \Sec{GW_sp_WKB}.
Then, in \Sec{sec:param}, we present some common parameterizations of the parameter $\alpM$
through the history of the universe that have been used in the literature,
and we present the numerical simulations that we perform to solve the GW equation using the {\sc Pencil Code} in \Sec{sec:numerical_solutions}.
In particular, we study the effects on the GW spectrum that can not be predicted using the WKB
In particular, we study the effects on the GW spectrum that cannot be predicted using the WKB
approximation.
Finally, we discuss potential observational implications of MG compared to GR
for the different parameterizations of $\alpM$ in \Sec{sec:obs_implications} and conclude in \Sec{sec:conclusions}.

Throughout the paper, we set $c = \hbar = \kB = 1$, use the metric signature $(-+++)$, and define the gravitational coupling constant
$\kappa = 8 \pi G_{\rm N}$, where $G_{\rm N}$ is the Newton constant.
We indicate with a prime
derivatives with respect to conformal time normalized by the
conformal Hubble rate at the time of GW generation $\eta \HH_*$ 
and with a dot derivatives with respect to cosmic time $t$.
Both times are related via the
scale factor as $a \dd \eta = \dd t$.

\section{Modified GWs on FLRW background}
\label{mod_GWs}

For simplicity, we assume the background evolution to follow $\Lam$CDM, while allowing a generic modification of the GW evolution.
Therefore, we start with
the homogeneous and isotropic background described by 
the Friedmann-Lema\^itre-Robertson-Walker (FLRW) metric.
Including tensor perturbations, the FLRW line element reads
\begin{equation}
\dd s^2 = a^2 \bigl(-\dd \eta^2 + [\delta_{ij} + a^{-1}
\hij (\xx, \eta)]\dd x^i\dd x^j\bigr),
\label{ds2}
\end{equation}
where $\hij = a\hij^\phys$ are
the strains obtained by scaling the physical strains $\hij^\phys$ with the scale factor $a$.
In GR, the GW equation in Fourier 
space\footnote{We use a tilde for variables in Fourier space and the following convention
\begin{equation}
    \tilde h (\kk) = \int h (\xx) \, e^{-i \kk \cdot \xx} \dd^3 \xx,
    \quad
    h (\xx) = \frac{1}{(2\pi)^3}\int \tilde h (\kk) \, e^{i\kk \cdot \xx} \dd^3 \kk.\nonumber
\end{equation}} 
reads \cite{Caprini:2018mtu,RoperPol:2018sap}
\begin{equation}
    {\tilde h}_\ij'' (\kk, \eta) +
    \biggl(k^2 - \frac{a''}{a} \biggr) \tilde h_{ij} 
    (\kk, \eta) = \frac{6}{a} \tilde T_{ij}^{\rm TT} (\kk, \eta),
    \label{GW_GR}
\end{equation}
where
$\tilde T_{ij}^{\rm TT} = \Lambda_{ijlm} \tilde T_{lm}$
is the traceless-transverse (TT) projection of the
normalized stress energy tensor, i.e., divided by the
radiation energy density $\EErad = 3H^2/\kappa$, being $H\equiv\dot a/a$ the Hubble rate at time $\eta$.
The projection operator is
$\Lambda_{ijlm} (\hat \kk) = P_{il} P_{jm} - \half P_{ij} P_{lm}$,
where $P_{ij} (\hat \kk) = \delta_{ij} - \hat k_i \hat k_j$ and
$\hat \kk = \kk/k$.
The scale factor at $\eta_*$ is set to unity,
and the wave numbers $k$ are also normalized by
the (normalized) conformal Hubble rate $\HH \equiv a'/a$.
We use an asterisk to refer to the time in the early universe
at which the GWs were generated.

Including the additional parameters $\alpM$ and $\alpT$ mentioned in \Sec{sec:introduction}, \Eq{GW_GR} generalizes 
to~\cite{Hwang:2001qk,Nunes+18}
\begin{equation}
\tilde h_{ij}'' (\kk, \eta) + \alpM \,\HH\, \tilde h_{ij}'(\kk, \eta)
+ \biggl(\cT^2 k^2 - \alpM \HH^2 -
\frac{a''}{a}\biggr)\, \tilde h_{ij}(\kk, \eta) = \frac{6}{a} \tilde
T_{ij}^{\rm TT} (\kk, \eta).
\label{eqn:GW_mod}
\end{equation}
The effect of massive gravitons could be included by adding a $\mg \neq 0$ term multiplying to $h_{ij}$ in \Eq{eqn:GW_mod}, as done in ref.~\cite{He:2021bqm}, but we omit it in the present work.

\EEq{eqn:GW_mod} can be recast in a simplified equation by introducing the new variable $\chi_\ij = e^{\ddd} \hij$~\cite{Belgacem:2017ihm,Belgacem:2018lbp,LISACosmologyWorkingGroup:2019mwx},
with
\begin{equation}
\ddd (\eta) = \frac{1}{2}\int_1^{\eta \HH_*} \alpM\, \hhh\dd \tau,
\end{equation}
where the integral is performed over normalized time $\tau = \eta'\, \HH_*$.
This yields
\begin{equation}
\tilde\chi_\ij''(\kk,\eta) + \biggl(\cT^2k^2 - \frac{\tilde a''}{\tilde a}\biggr)\tilde\chi_\ij(\kk,\eta) = \frac{6}{a}\tilde T_\ij^{\rm TT}(\kk,\eta),
\label{eqn:GW_mod_chi}
\end{equation}
where $\tilde a = a e^{\ddd}$,
and $\tilde a''/\tilde a$ can be assembled from $a$ and  $\alpM$ as
\begin{equation}
\frac{\tilde a''}{\tilde a} = \frac{a''}{a} \biggl(1 + \frac{1}{2}
\alpM \biggr) + \frac{1}{2} \alpM \HH^2 \biggl(1 + \frac{1}{2}
\alpM \biggr) + \frac{1}{2} \alpM' \HH.
\label{eqn:tilde_app_a}
\end{equation}

The variable $\chi$ is related to the difference between GW and EM luminosity distances mentioned in \Sec{sec:introduction}.
We first express the latter in $\Lam$CDM,
\begin{equation}
\dEM
= \frac{a_0}{a H_0} \int_a^{a_0} \frac{a_0 \dd a'}{a'^2 \sqrt{\Om(a)}},
\label{eqn:dEM_z_a}
\end{equation}
where $H_0$ is today's Hubble rate
and $\Omega(a)$ is the total energy density, which includes matter,
radiation, and dark energy contributions; see 
\App{sec:appendix_Friedmann}.
The GW luminosity distance is
\begin{equation}
\dGW = \frac{a}{\tilde a}\dEM = e^{-\ddd}\dEM,
\label{eqn:dGW_dEM}
\end{equation}
which reduces to $\dGW = \dEM$ in GR; see ref.~\cite{Tasinato:2021wol} for a detailed calculation.

Previous numerical work was usually restricted to specific eras, e.g.,
radiation (RD) \cite{RoperPol:2018sap, RoperPol:2019wvy, Kahniashvili:2020jgm,
Brandenburg:2021bvg, RoperPol:2021xnd, Brandenburg:2021tmp,
RoperPol:2022iel, Sharma:2022ysf} or matter domination (MD)
\cite{Brandenburg:2021pdv, Brandenburg:2021bfx, He:2021kah}, using
a constant equation of state (EOS), defined to be $w\equiv p/\rho$,
being $p$ the pressure and $\rho$ the energy density, such that
$w = 1/3$ and $0$ during RD and MD, respectively.
The dynamical thermal history of the universe during RD,
represented by the relativistic $g_*$ and adiabatic $\gS$ degrees of freedom,
was also ignored.
In GR, this formulation is justified since the evolution
of the physical strains $h_\ij^\phys$ when the source is inactive ($T_\ij \sim 0$) can
be approximated to dilute as $h_\ij^{\rm phys}
\propto a^{-1}$ if one
neglects the evolving relativistic degrees of freedom and
transitions between radiation, matter, and dark energy 
dominations ($\Lam$D) \cite{Saikawa:2018rcs}.
In the present work, we focus on including such effects
and solve the GW equation in modified gravity, 
which present richer dynamics even when
the source is inactive, from the time of GW generation
up to present time.

Assuming a piece-wise EOS,
such that $w = 1/3$, $0$, and $-1$ during RD, MD, and 
$\Lam$D, respectively,
would create discontinuities in the time evolution of $a(\eta)$ and
its derivatives that appear in \Eqs{eqn:GW_mod}{eqn:GW_mod_chi}.
Therefore, 
to find a smooth $a(\eta)$,
we directly solve the Friedmann equations (see \App{sec:appendix_Friedmann} for details),
\begin{equation}
   \HH = \frac{H_0}{\HH_*} \, a \sqrt{\Omega (a)},
    \quad 
    \frac{a''}{a} = \frac{1}{2} \HH^2
    \bigl[1 - 3 w (a) \bigr],
    \label{eqn:a_derivatives_omega}
\end{equation}
where $\Om(a)$ denotes the total energy density,
and the term $\HH_*$ appears due to our definition of the
normalized $\HH = a'/a$.
We set $a_* = 1$ for consistency
with our GW equation; see the discussion below \Eq{GW_GR}.
Using \Eq{eqn:a_derivatives_omega}, we solve for the time evolution of the scale factor
$a(\eta)$ and finally obtain the time evolution of
$\HH$ and $a''$ that are required to solve the GW equation, either when they appear explicitly in \Eq{eqn:GW_mod} for $h_\ij$ or to compute the term $\tilde a''$, defined in \Eq{eqn:tilde_app_a}, that appears in \Eq{eqn:GW_mod_chi} for $\chi_\ij$.

\section{WKB approximation}
\label{WKB_s}

\subsection{Solution of the GW equation}
\label{sec:WKB}

Modifications of the GW propagation,
in the absence of sources\footnote{%
In the absence of sources, the GW equation [see \Eq{eqn:GW_mod}] does
not depend on the wave vector $\kk$ but only on its modulus (the wave number) $k$.
Hence, the solution can simply be expressed as a function
of $k$.}
(i.e., $T_{ij} = 0$),
have been studied using the WKB
approximation \cite{Nishizawa17,Arai+17}.
The WKB solution can be obtained using the following ansatz
\begin{equation}
\tilde\chi_\ij(k, \eta) = \chi(k, \eta)\,e_\ij = Ae_\ij\, e^{iB},
\label{ansatz_WKB}    
\end{equation}
where $A$ and $B$ are generic coefficients, and $e_{ij}$ is the polarization tensor.\footnote{
In the absence of sources, the GW propagation of any
polarization mode is the same, so we can just call $\chi$
the amplitude of each mode.
If the produced GW signal is polarized, then for each mode we
need to impose the corresponding initial conditions.}
Substituting this into \Eq{eqn:GW_mod_chi}, one gets
\begin{align}
0 & = 2\frac{A'}{A} + \frac{B''}{B'} 
\quad\quad\quad\quad\quad\quad
\Map A(\eta) = e^{-\tilde\ddd},
\label{eq1_WKB}\\
B'^2 & = \frac{A''}{A} + k^2\cT^2 - \frac{\tilde a''}{\tilde a}\approx k^2\cT^2
\Map B(k,\eta) = \pm k(\eta\hhh_* - \Del T),
\label{eq2_WKB}
\end{align}
where we have neglected $A''/A$ and $\tilde a''/\tilde a$ compared to $k^2 \cT^2$, following the WKB approximation.
We have defined the
additional damping
factor $\tilde\ddd$, and the time delay due to the effective
GW speed $\Del T$
\begin{equation}
\tilde\ddd = \frac{1}{2}\int_1^{\eta\hhh_*}\frac{B''}
{B'}\dd \tau = \frac{1}
{2}\int_1^{\eta\hhh_*}\frac{\cT'}
{\cT}\dd \tau, \quad
\Del T = \int_1^{\eta\hhh_*}(1 - \cT)\dd\tau.
\label{eqn:ddd_DelT}
\end{equation}
Hence, the particular solution is
\begin{equation}
\chi(k,\eta) = e^{-\tilde\ddd}e^{\pm ik(\eta\hhh_* - \Del T)} = e^{-\tilde\ddd\mp ik\Del T}\chi^\GR(k,\eta).
\label{h_part_sol}
\end{equation}
where $\tilde\ddd = \Delta T = 0$ in GR.
For the initial conditions\footnote{In the current work, we focus on the propagation of a GW background after it has already
been generated and reached a stationary solution, when the source
is no longer active.
Since the propagation only depends on $k$, the initial conditions
can be computed from the 3D fields in Fourier space, after
shell integration over directions $\hat \kk$, for each polarization mode.
}
$\chi(k, \eta_*) = \chi_* (k)$ and $\chi' (k, \eta_*) = \chi_*' (k)$, 
the general solution reads
\begin{align}
\chi(k, \eta) & = e^{-\tilde\ddd}\Bigg[\chi_*(k)\cos k\tilde c_{\rm T} (\eta\hhh_* - 1 ) + \frac{\cT^* \chi_*'(k) + \half {\cT'}^* \, \chi_*(k)}{k {\cT^*}^2}\sin k\tilde c_{\rm T}
(\eta\hhh_* - 1)\Bigg],
\label{chi_modGR}
\end{align}
where $\tilde c_{\rm T}
(\eta) = \int \cT \dd \eta'/(\eta \HH_* - 1)$, and $\cT^*$ and ${\cT'}^*$ are the values of the GW
speed (and its derivative) at the initial time of GW 
production.
Note that the specific choice of initial conditions
does not allow to give a linear relation between
$\chi$ and $\chi^\GR$
as done in ref.~\cite{Nishizawa17} and, in general,
$\chi\neq e^{-\tilde\ddd - i k \Delta T}\chi^\GR$, 
which is only true
when referring to the particular solution of the ODE, given
in \Eq{h_part_sol}.
In terms of the initial conditions $h (k, \eta_*) = h_* (k) = \chi_* (k)$ and 
$h'(k, \eta_*) = h'_* (k) = \chi'_*(k) - \half \alpM^*\,
\chi_*(k)$,
the solution \eqref{chi_modGR} can be rewritten as
\begin{align}
    \chi (k, \eta) = e^{-\tilde \ddd}
    \biggl[&\,h_*(k) \cos k \tilde c_{\rm T} (\eta \HH_* - 1)
    \nonumber \\ &\, +
    \frac{\cT^* h_*'(k) + \half \bigl(\cT^*\, \alpM^* + {\cT'}^*
    \bigr) h_* (k)}{ k {\cT^*}^2} \sin k \tilde c_{\rm T}
    (\eta \HH_* - 1) \biggr].
    \label{eq_chi_modGW_hij}
\end{align}
The general solution reduces to the following when $\cT = 1$
at all times,
\begin{equation}
\chi(k,\eta) = h_*(k)\cos k(\eta\hhh_* - 1 ) + \frac{h_*'(k) + \half \alpM^* \, h_* (k)}{k}\sin k(\eta\hhh_* - 1).
\label{simp_WKB}
\end{equation}
This expression holds both in GR with $h(k, \eta) = \chi(k, \eta)$ and
$\alpM = 0$,
and in MG theories with $h(k, \eta) = e^{-\ddd}
\chi (k, \eta)$ and $\alpM \neq 0$.

\subsection{Limitations of the WKB approximation}
\label{limitations_WKB}

To obtain the WKB solution, we have neglected two terms in \Eq{eq2_WKB}, which can be expressed as
\begin{align}
\frac{A''}{A} - \frac{\tilde a''}{\tilde a} 
& = 
- \frac{\cT''}{2\cT}
+ \frac{3\cT'^2}{4\cT^2} 
- \frac{1}{2}\alpM\Big(1 + \frac{1}{2}\alpM\Big)\hhh^2 
- \frac{1}{2}\alpM'\hhh 
- \frac{a''}{a}\Big(1 + \frac{1}{2}\alpM\Big),
\label{eqn:limits_all}
\end{align}
where we have used $\HH' = a''/a - \HH^2$.
Hence, the WKB assumption could break down when at least one
of these terms is not negligible when compared to $k^2 \cT^2$.
First, we obtain the limiting values of $k$ due to $\cT$ and its derivatives in the first two terms on the right-hand side of \Eq{eqn:limits_all}
\begin{equation}
k_{\rm lim,\cT''}\lesssim\sqrt{\frac{|\cT''|}{2\cT^3}},\quad
k_{\rm lim,\cT'}\lesssim\frac{\sqrt{3}|\cT'|}{2\cT^2}.
\label{WKB_limitations_cT}
\end{equation}
Note that in this paper we only consider $\cT$ to be constant in time, 
i.e., $k_{\rm lim,\cT''}$ and $k_{\rm lim,\cT'}$ are zero.
We defer the study of time-varying $\cT$ and its effects to future work.

In addition,
we have one term that depends on $\alpM$, one term that
depends on the time evolution of $\alpM$, and one term
that depends on $a''/a$.
The latter appears also in GR when WKB is used to approximate the
solution for the scaled strains, which do not decay (note that
the decay of the physical strains is already absorbed by the
scale factor), as shown in \Eq{simp_WKB}.
This term is given in \Eq{eqn:a_derivatives_omega} and its upper
bound can be found using Friedmann equations
\begin{equation}
    \frac{a''}{a} \lesssim \frac{1}{2}
    \frac{\Om_{\mat, 0}}{\Om_{\rad, 0}}
    \frac{g_*}{g_*^0}
    \biggl(\frac{\gS}{\gS^0}\biggr)^{-{4\over3}}
    \HH \, \frac{a_*}{a_0} \lesssim 
    10^4 \, h^2 \, \HH \, \frac{a_*}{a_0},
    \label{rel_app}
\end{equation}
where the first inequality is an approximation valid during 
the RD era and it decays during MD.
Hence, the upper bound is valid at all times and its specific value depends
on $a_*/a_0$.
For example, for GWs generated at the EWPT or at the QCDPT, this leads to the
following limiting wave numbers (for $h=0.67$), at which the WKB approximation might
break down,
\begin{align}
    &\, \klimap^\EW \sim 1.87 \times 10^{-6} \, \HH \,
    \sqrt{\frac{\bigl|1 + \half \alpM\bigr|}{\cT}} \leq
    1.87 \times 10^{-6} \, \sqrt{\frac{\bigl|1 + \half \alpM\bigr|}{\cT}},
    \label{kcritEW}\\
    &\, \klimap^\QCD \sim 5.90 \times 10^{-5} \, \HH \,
    \sqrt{\frac{\bigl|1 + \half \alpM\bigr|}{\cT}} \leq
    5.90 \times 10^{-5} \, \sqrt{\frac{\bigl|1 + \half \alpM\bigr|}{\cT}}.
    \label{kcritQCD}
\end{align}
In both cases, this term is subdominant up to very large
superhorizon scales and it is bounded by the values in \Eqs{kcritEW}{kcritQCD} since $\HH = \eta_*/\eta$
during RD era as can be seen using \Eq{eqn:a_derivatives_omega}.
This limit can be modified by the inclusion of $\alpM$ and $\alpT$
but unless they take large values these modifications are negligible.

On the other hand, two additional limitations to
the WKB approximation appear due to the $\alpM$ parameter,
\begin{align}
    \klimaM &\, \sim \frac{\HH}{\cT}\sqrt{\biggl|\frac{\alpM}{2} \Bigl(1 + \half \alpM 
    \Bigr)\biggr|} \lesssim \frac{1}{\cT}\sqrt{\frac{|\alpM^*|}{2}},
    \label{WKB_limitations0} \\
    \klimaMp &\, \sim \frac{1}{\cT}\sqrt{\frac{\HH \, |\alpM'|}{2}} \leq
    \frac{1}{\cT}\sqrt{\frac{|\alpha^{'*}_{\rm M}|}{2}},
    \label{WKB_limitations}
\end{align}
where we can neglect the term $\half \alpM$ in front of 1 in the
first limit
for small values of $\alpM$.
Hence, the WKB limit can break down in MG
around the horizon or at larger scales.
In general, we expect the limit from $\alpM$ to be more
restrictive than that from $\alpM'$ and to dominate
at the initial time when $\HH = 1$.
However, this depends on the parameterization of $\alpM$,
which can give different results for both limits; see \Sec{sec:param}.

\subsection{GW spectrum using the WKB approximation}
\label{GW_sp_WKB}

The spectrum of GW energy density can be expressed as
\begin{equation}
\OmGW (k, \eta) = \frac{1}{\rho_{\crit,0}}
\frac{\dd \rho_\GW}
{\dd \ln k} = \frac{1}{6} \biggl(\frac{H_*}{H_0} \biggr)^2 \biggl(
\frac{a_*}{a_0} \biggr)^4 k S_{h'} (k, \eta),
\label{eqn:OmGW_Sh}
\end{equation}
where $2 S_{h'}$ is the spectrum\footnote{
Following ref.~\cite{RoperPol:2018sap}, we define the spectrum
$S_{h'} (k, \eta)$ from the $+$ and $\times$ polarization
modes, giving an extra factor of 2 due to the property
\begin{equation}
     \tilde h'_{ij} (\kk, \eta) \, \tilde h^{'*}_{ij} (\kk', \eta)
    = 2 \bigl[h'_+ (\kk, \eta) \, h^{'*}_+ (\kk', \eta) + 
    h'_\times (\kk, \eta) \, h^{'*}_\times (\kk', \eta) \bigr]. \nonumber
\end{equation}}
of the scaled strain
derivatives,
\begin{equation}
    \bra{\tilde h_{ij}' (\kk, \eta) \, \tilde h_\ij^{'*} (\kk', \eta)} = (2 \pi)^6 
    \delta^3 (\kk - \kk') \frac{2 S_{h'} (k, \eta)}{4 \pi k ^2}.
\label{eqn:Sh}
\end{equation}
Taking $\ddd' = \frac{1}{2}\alpM\hhh$,
the modified GW energy spectrum can be obtained using the WKB
solution for MG, given in \Eq{chi_modGR},
and assuming that at time $\eta_*$ 
the GWs are in the free-propagation 
regime following GR, 
such that $|h'_*(k)| = k \, |h_*(k)|$,
\begin{align}
   \OmGW &\, (k, \eta) = e^{-2\ddd} \OmGW^\GR (k, \eta)\, \xi (k, \eta),
   \label{OmGW_mod_general}
\end{align}
where $\xi (k, \eta)$ corresponds to the spectral modifications with
respect to the GR spectrum after taking into account the amplification
(or depletion) produced by the damping factor $\ddd$ and assuming constant-in-time $\cT$
(see \App{sec:WKB_spec} for details and, in particular, \Eq{general_xi_modGR} for generic time-dependent $\cT$),
\begin{equation}
    \xi (k, \eta) = 1 + \frac{1}{2}\alpT
    + \frac{\alpM^*}{2 k}
    + \frac{1}{8 k^2} \Biggl[
    \alpM^2\HH^2\biggl(1 + \frac{1}{\cT^2}\biggr) 
    + \alpM^{*2}
    \Biggr]
    + \frac{\alpM^*\alpM^2\HH^2}{8 k^3\cT^2}
    + \frac{\alpM^{*2}\alpM^2 \HH^2}{32 k^4 \cT^2}.
\label{eqn:EGW_mod}
\end{equation}
Note that we have averaged over oscillations in time $\eta$,
such that the expression is valid at $k \, \tilde c_{\rm T} \gtrsim (\eta \HH_* - 1)^{-1}$.
At late times, $\HH^2 = (\eta_*/\eta)^2$ becomes small, so we can neglect the
terms that dilute with $\HH^2$,
\begin{equation}
    \xi (k, \eta \gg \eta_*) = \half \bigl(1 + \alpT\bigr)
    + \half \Bigl(1 + \frac{\alpM^*}{2k} \Bigr)^2.
    \label{xi_mod_general}
\end{equation}
From \Eq{xi_mod_general}, we can identify the IR and UV
limiting ranges of the spectrum,
\begin{align}
\OmGW^\IR = \OmGW \Bigl(k \ll \half \alpM^*, \eta \gg \eta_* \Bigr)
= & \, \frac{{\alpM^*}^2}{8 k^2} \, e^{-2\ddd} \, \OmGW^\GR(k,\eta),
\label{OmGW_mod_IR}\\
\OmGW^\UV = \OmGW \Bigl(k \gg \half \alpM^*, \eta \gg \eta_*\Bigr)
= & \, \Bigl(1 + \half \alpT\Bigr) \, e^{-2\ddd}  \, \OmGW^\GR(k,\eta).
\label{GW_UV_mod}
\end{align}
Note that the factor ${\alpM^*}^2$ is part of the IR limit,
indicating that the IR enhancement of the form $k^{-2}$ holds 
regardless of the sign of $\alpM^*$.
The critical $k_\crit = \half |\alpM^*|$ indicates where the IR regime begins, 
i.e., where the $k^{-2}$ term in \Eq{xi_mod_general} becomes dominant.

\EEq{eqn:EGW_mod} shows that the GW spectrum in the IR regime can
present up to $k^{-4}$ modifications to the
GR spectrum but this and other terms vanish as time
evolves since they are proportional to $\HH^2$.
Hence, at late times, we end up with the GR spectrum
amplified (or diluted) for negative (or positive) $\ddd$ by $e^{-2\ddd}$ at all wave numbers
with a $k^{-2}$ enhancement in the IR regime,
proportional to the additional $\eigth {\alpM^*}^2$
factor, as shown in \Eq{OmGW_mod_IR}.
In the UV regime, we find the $e^{-2\ddd}$ enhancement (or depletion) and an
additional factor $1 + \half \alpT$, which is, in general, a function of $\eta$.
The sign of $\alpT$ determines if the GW spectrum is 
amplified or decreased with respect to that obtained from
GR.

We have found that, according to the WKB approximation,
the parameter $\alpM$ introduces changes in the spectral shape
at $k \leq k_\crit = \half |\alpM^*|$ that do not dilute as time
evolves and depend on the value of $\alpM$ only at the time of
GW generation.
On the other hand, note that, using \Eqs{WKB_limitations0}{WKB_limitations}, we can estimate 
the WKB approximation to break down around
$k \sim \max(\klimaM, \klimaMp)$.
This means that when $\HH \sim 1$, the spectral changes occur at wave numbers around and below the
critical $k_\crit$,
where the WKB approximation might not be valid (for $\cT \sim 1$).

We investigate the resulting spectra in MG using
numerical simulations in \Sec{sec:numerical_solutions}
with the objective to test the validity of the WKB
approximation and its potential limitations and to confirm the resulting GW spectra
when MG parameters are introduced.

\section{Phenomenological parameterizations}
\label{sec:param}

In \Sec{GW_sp_WKB}, we noted that, according to the WKB 
approximation, a constant $\alpT \neq 0$ induces modifications in the total GW energy density, but not
in its spectral shape as long as $\alpT \ll 1$.
Realistically, the tensor speed excess at the present day is observationally constrained to be $\alpTz\lesssim\ooo(10^{-15})$ 
by the binary neutron star merger GW170817 and its gamma-ray burst GRB 170817A~\cite{LIGOScientific:2017zic}.
This constraint can be circumvented if $\alpT$ is either frequency- or time-dependent, such that larger deviations could hide outside the LIGO--Virgo frequency 
band~\cite{deRham:2018red,LISACosmologyWorkingGroup:2022wjo}
or in the past~\cite{Cai:2015yza,Cai:2016ldn}.
In this study, we show that even using larger values of
$\alpT$, constant in time and frequency, the modifications to 
the GW spectrum are negligible, which is seen from the WKB
approximation and can be confirmed with numerical simulations.
Hence, from now on, we will focus on two cases: 
{\em(i)} 
constant $\alpT$ with $\alpM = 0$, 
and {\em(ii)} $\alpT = 0$ allowing $\alpM$ to take
different values as a function of time.

In particular, we are interested in the perspectives of negative values of $\alpM$, 
as they would enhance the amplitude of the GW spectrum.
Various constraints on the present-day value $\alpMz$ exist in
the literature,
but are significantly less stringent than those on $\alpT$.
Assuming the cosmic acceleration due to scalar-tensor theories
of gravity, it can be shown that $|\alpM| \lesssim 0.5$ at 
late times (see figure~4 in ref.~\cite{Lombriser+15}).
This is compatible with the numerical results of ref.~\cite{Arai+17} (see their figure~2), giving $|\alpMz| \lesssim 1$.
On the other hand, the Planck Collaboration reports
a slightly tighter constraint on $\alpMz \gtrsim -0.1$
(see table 8 of ref.~\cite{Planck:2018vyg}).
Note that they parameterize $\alpM = \alpMz \, (a/a_0)^n$ with
$n \in (0.5, 1)$.
In addition, to avoid having an infinite number of GW sources in the early Universe, 
$\dGW$ is required to increase monotonically at earlier times,
i.e., $\dd(\dGW)/\dd a\leq0$, as argued in ref.~\cite{Mastrogiovanni:2020gua}.
Hence,
using \Eqss{eqn:dEM_z_a}{eqn:dGW_dEM}, one gets a bound for $\alpM$,
\begin{equation}
\alpM(a)\geq\frac{a_0}{a\sqrt{\Om(a)}}\Bigg[\int_a^{a_0}\frac{a_0 \, da'}{a'^2\sqrt{\Om(a')}}\Bigg]^{-1} - 1,
\label{eqn:alpM_bound}
\end{equation}
which is satisfied by all positive values
of $\alpM$ and for negative values with $|\alpM|\leq|\alpM(a_*)|\lesssim 1$.
Taking these different constraints into account,
we explore a range of $\alpMz\in[-0.5, 0.3]$.

Besides the simplest consideration of a constant $\alpM$ in time, 
which we call choice 0,
its time-dependent forms can be written in accordance with
specific gravity models~\cite{Odintsov+22}.
Following refs.~\cite{Bellini+14,Nunes+18,Kennedy:2018gtx}, we choose phenomenological forms of $\alpM$
as simplified parameterizations motivated by effective descriptions
of modified gravity~\cite{Gleyzes:2017kpi}.
Specifically:
\begin{equation}
\alpM(\eta) = 
\begin{dcases}
\alpMz\biggl[\frac{a(\eta)}{a_0}\biggr]^n & (\text{choice I}),\\
\alpMz \frac{1}{\Om(\eta)} & (\text{choice II}),\\
\alpMz \frac{1-\Om_\mat(\eta)/\Om(\eta)}{1-\Om_{\mat,0}} & (\text{choice III}),
\end{dcases}
\label{eqn:alpM_param}
\end{equation}
where choices II and III give a value of $\alpM$ proportional
to the percentage of dark energy density
and the combination of dark and radiation
energy densities, respectively, at each time $\eta$ compared to
their relative amounts at the present time.
The exponent $n$ in choice I needs to satisfy the following stability conditions
\cite{Denissenya:2018mqs}:
\begin{equation}
\mbox{stability for:}\;
\begin{dcases}
0 < n < \threehalf \, \Om_{\mat,0} \simeq \half, \ & \mbox{if}\; \alpMz > 0, \\
n > \threehalf, \ & \mbox{if}\; \alpMz < 0.
\end{dcases}
\label{stab_cond}
\end{equation}

\begin{figure}[t]
\centering
\includegraphics[width=0.495\textwidth]{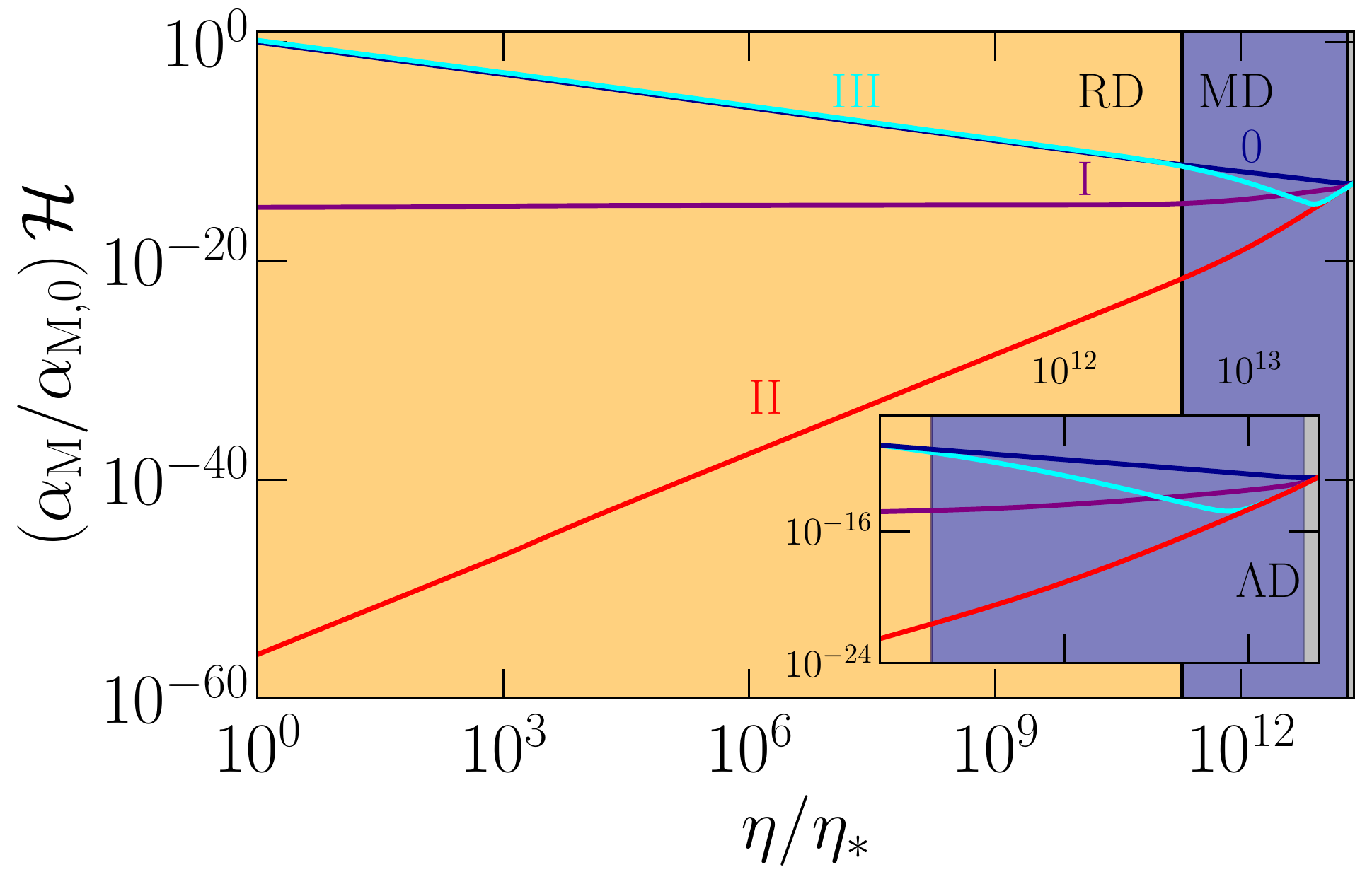}
\includegraphics[width=0.465\textwidth]{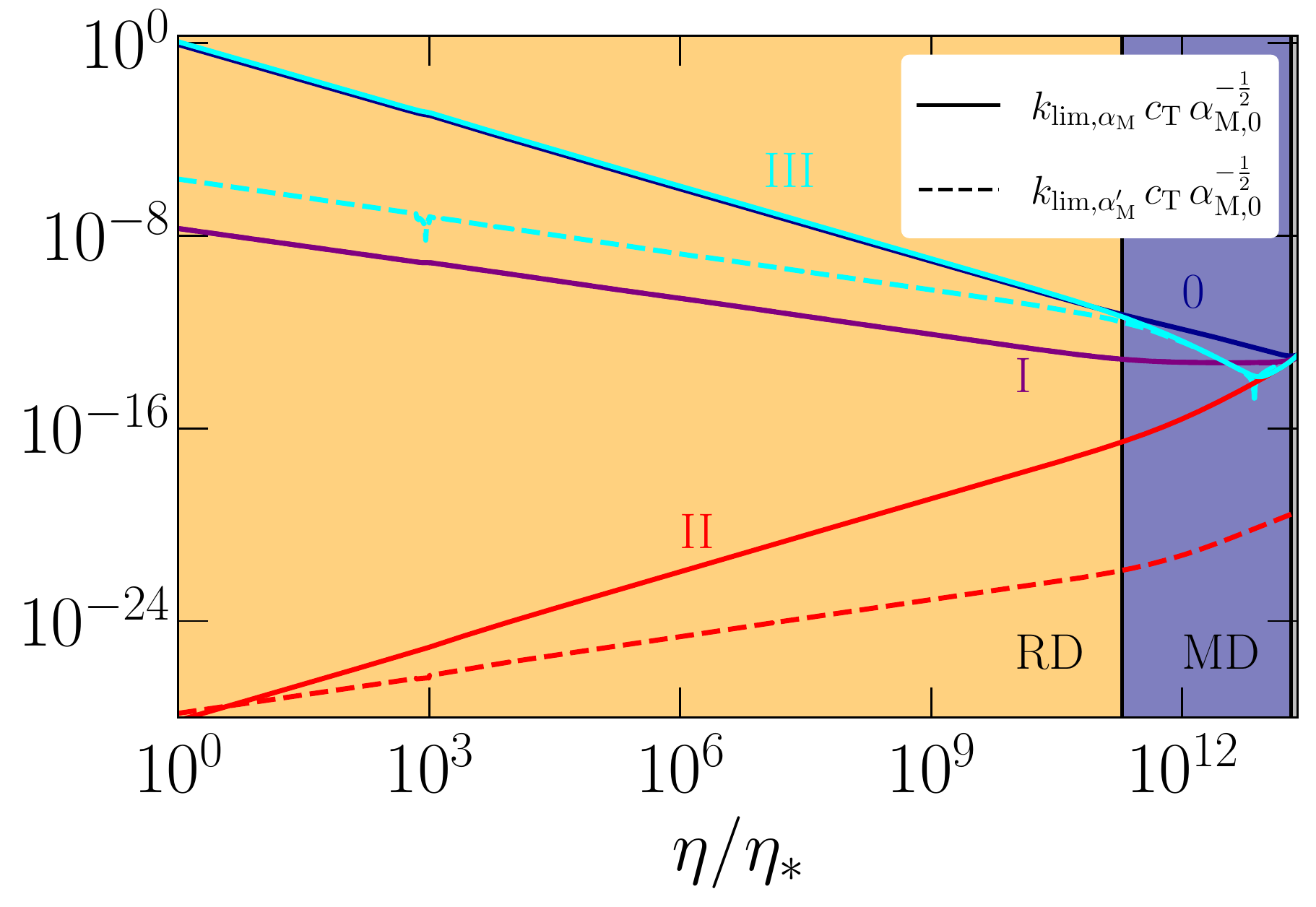}
\caption{
Time evolution of $(\alpM/\alpMz)\hhh$ (left panel), which
contributes to the change of amplitude over time,
$h_{ij} \sim e^{-\ddd}$ with $\ddd = \half \int^\eta \alpM \HH \dd \eta'$,
and of $\klimaM \, \cT$ and $\klimaMp \, \cT$ (right panel), which
are the
terms neglected when compared to $k \cT$ under the WKB approximation;
see \Eqs{WKB_limitations0}{WKB_limitations}.
All four parameterization choices (0 to III) are shown and,
for illustrative purposes, $n = 1$ is chosen.
The main figures show the full evolution whereas
the inset (in left panel) shows only times after the onset of MD.
We have taken $\eta_*$ to correspond to the EWPT for the specific
values in the axes, which puts the present time at
$\eta_0/\eta_* \simeq 2.4 \times 10^{13}$.
}
\label{fig:alpM_Hcal}
\end{figure}

To provide an intuition on the aforementioned $\alpM$ parameterizations,
we show in \Fig{fig:alpM_Hcal} (left panel), for the different time dependencies
given in \Eq{eqn:alpM_param}, the evolution of $\alpM\hhh$, which
characterizes the growth or damping produced by $\alpM$, according to
the WKB approximation [see \Eq{eqn:ddd_DelT}].
We also show the time evolutions of $\klimaM \, \cT$ and
$\klimaMp \, \cT$ (right panel), which correspond to the wave numbers
at which the WKB approximation might break down,
according to \Eqs{WKB_limitations0}{WKB_limitations}.
\FFig{fig:alpM_Hcal} shows the time evolution from the time
of generation (e.g., EWPT) up to present time.
We see that for choices 0 and III, $\alpM\hhh$ converges to the same values during RD
and $\Lam$D, and only becomes different during MD, as expected.
On the other hand, for choices I and II, $\alpM \HH$ is negligible
for all of RD and most of MD and rapidly increases later on, especially during $\Lam$D.
For all parameterizations, the GW friction converges to the value $\alpMz$ at present time.

We see in \Fig{fig:alpM_Hcal} (right panel) that for choices 0 and III,
the limiting $k$ is around the horizon scale at the
time of generation, i.e., $\klimaM \, \cT \sim \sqrt{\alpM^*/2}$
and then it decreases into superhorizon scales since it
is proportional to $\HH = \eta_*/\eta$ (this is because
for these choices, $\alpM \simeq \alpM^*$ during the RD era).
The term $\alpM' = 0$ for choice 0 and $\alpM' \ll \alpM$ for choice III, as can be
seen in \Fig{fig:alpM_Hcal}.
The ratio between $\alpM'$ and $\alpM$ at the time of GW generation, within RD,
can be found to be (see the associated online material \cite{GH})
\begin{equation}
    \biggl(\frac{|\alpM'|}{\alpM} \biggr)^{1\over2}_{\rm III} \lesssim 
    \biggl[\frac{\Omega_{\mat, 0}}{\Omega_{\rad, 0}} \frac{g_*}{g_*^0} \biggl(
    \frac{\gS}{\gS^0}\biggr)^{-{4\over3}} \frac{a_*}{a_0}\biggr]^{1\over2}
    \lesssim
    \sqrt{2} 
    \times 10^2 \, h
    \biggl(\frac{a_*}{a_0}\biggr)^{1\over2},
    \label{rel_alpMp}
\end{equation}
which only becomes of order $1$ when the time of GW generation is around 
the end of the RD era (see right panel of \Fig{fig:alpM_Hcal}).
For choices I and II, although the two terms are of the 
same order, ${\alpha}^{'*}_{\rm M} \sim \alpM^*$, their value
at the time of GW generation is much smaller than
$\alpMz$ (for all $n > 0$ in the choice I).
Therefore, for choices 0 and III, the WKB estimate might break down around the horizon and at superhorizon scales, while for choices I and II, the approximation is estimated to be valid also at superhorizon scales (down to the limiting wave number).
Hence, for the latter, we do not expect to observe
any relevant spectral change, since the IR enhancement
is determined by the value of $\alpM^*$, as seen in \Eq{xi_mod_general}.

Note that, since many efforts of modifying gravity are aimed at
addressing the late-time acceleration of the universe, the
$\alpM$ parameterization choices are commonly constructed to be dominant in the late universe.
This is indeed the case for choices I and II here.
If one considers only the MD and $\Lam$D history of the universe,
choice III is also well-motivated to be relevant at late times.
However, here we explore the entire history of the universe from well within RD,
which means that $\alpM\hhh$ becomes dominant for choice III both at early and late times.
We note that the physical motivations of the choices here can be potentially ambiguous.
Even though there exists a wide range of discussions on MG during inflation~\cite{Nunes+18,Yoshimura:2022kyd} and around recombination~\cite{Zumalacarregui:2020cjh,Braglia:2020auw},
where $\alpM$ is essentially a free function of the scalar field,
there seems to be a 
relative lack of numerical studies on the effects of $\alpM$
during RD
(note, however, a brief discussion of $\alpM = -1$ during RD in ref.~\cite{Linder:2021pek}).
Hence,
we emphasize the aim of this work to provide an
understanding of the phenomenological behavior of the GW spectrum due to $\alpM$ 
for signals produced during RD, although the GW background computed in \Sec{GW_sp_WKB} for MG is relevant in the most general case.

\section{Numerical solutions}
\label{sec:numerical_solutions}

To explore the limits and validity of the WKB approximation, we use
the {\sc Pencil Code} to numerically solve the GW equation under
MG, given in \Eq{eqn:GW_mod}.
The {\sc Pencil Code} is a highly parallelized modular code that can be
used to solve various differential equations \cite{pencil}.
In the context of cosmological GWs, it has previously been used to study GWs generated by hydrodynamic and
MHD stresses in the early universe \cite{RoperPol:2019wvy}.
It uses a GW solver that advances the strains at each time step
sourced by the anisotropic stresses that are separately computed as the
solution to the MHD equations \cite{RoperPol:2018sap}.
Previous numerical work solved the GW equation under GR while, in the
present work, we have extended the code to solve \Eq{eqn:GW_mod}; see
\App{sec:numerical_scheme} for further details on the numerical scheme.

\subsection{Initial condition and time stepping schemes}
\label{init_conds}

\begin{figure}[t]
\centering
\includegraphics[width=.502\textwidth]{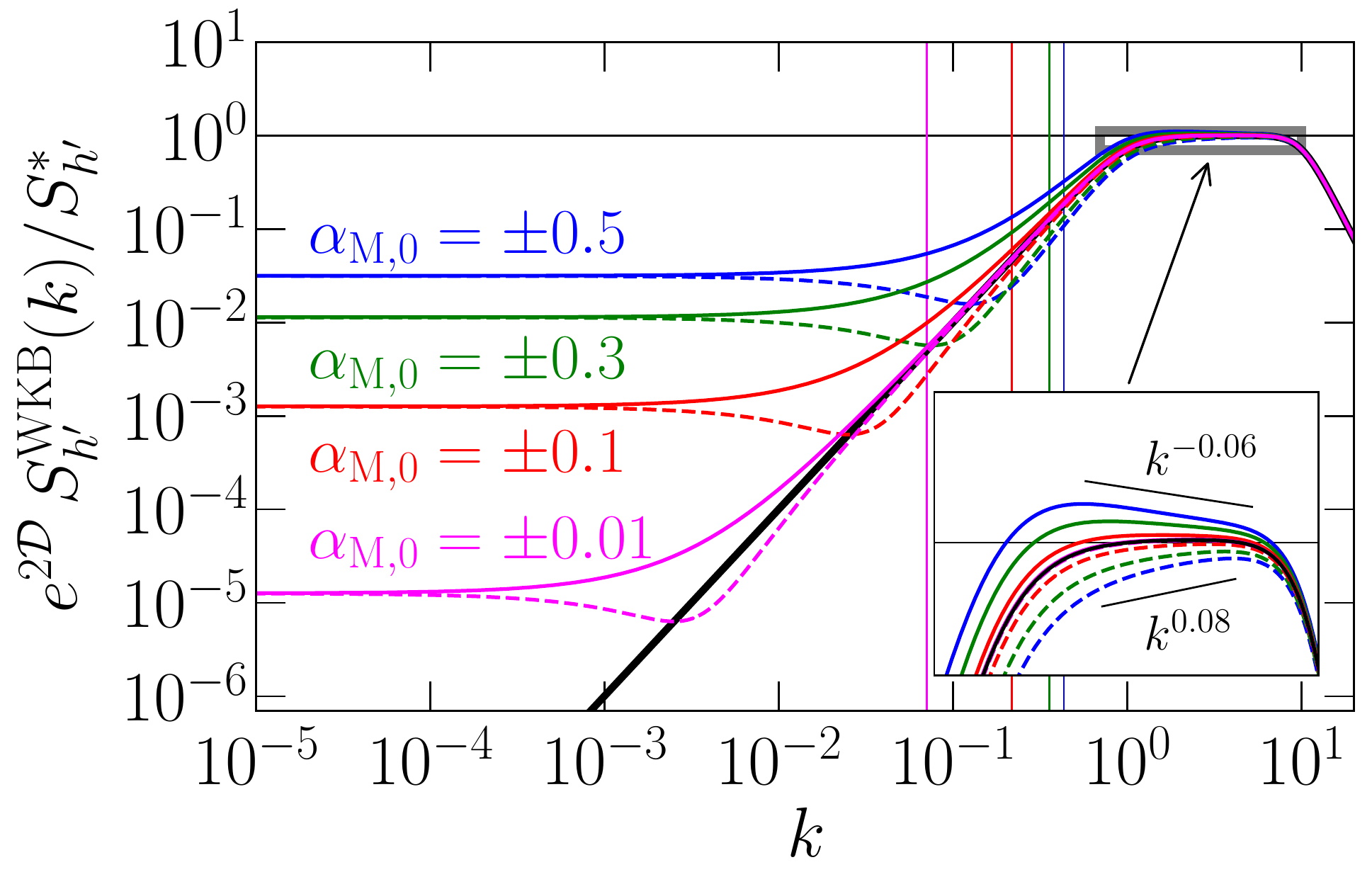}
\includegraphics[width=.49\textwidth]{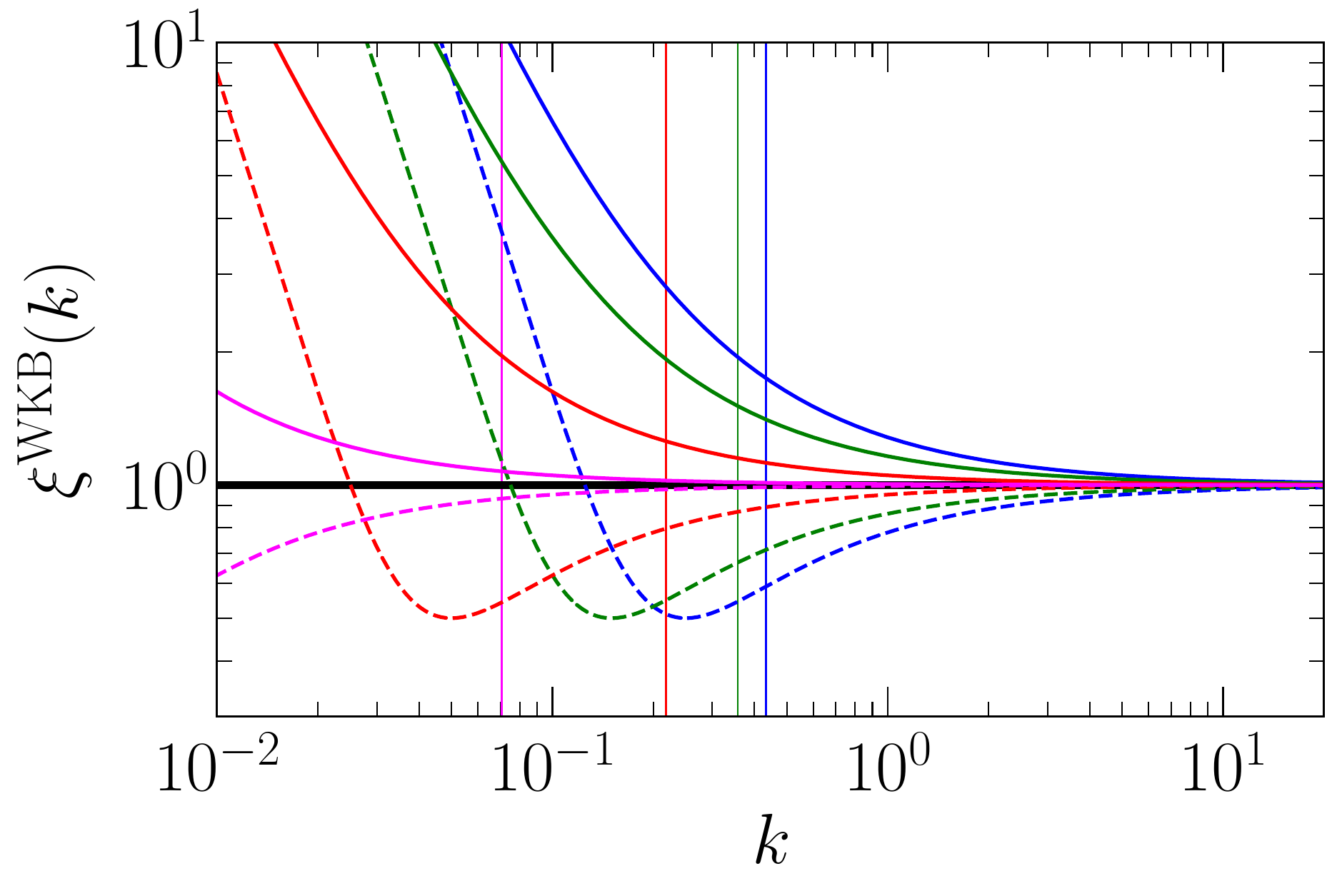}
\caption{{\em Left panel: }Expected WKB spectra [see \Eq{OmGW_mod_general}] of the strain derivatives $S_{h'} (k)$,
compensated by $e^{-2\ddd}$,
at late times due to non-zero $\alpha_{\rm M}^*$ at the time of generation for an initial GW background characterized by a double broken power law, given in \Eq{sp_ini}.
The horizontal line shows the value at the peak $S_{h'}^*$.
{\em Right panel: }Expected spectral modifications
$\xi^\WKB (k) \equiv e^{2\ddd} S_{h'}^\WKB (k)/S_{h'}^\GR (k)$,
given in \Eq{xi_mod_general}.
The black curves indicate the GR solution and the colored curves 
correspond to different values of $\alpM^*$, with the solid and
dashed ones being positive and negative values, respectively.
Vertical lines indicate 
the estimated value of $\klimaM$, below which the WKB approximation
might fail.
The inset zooms
in on the behaviors around the horizon at $k \in (1, 10)$.}
\label{fig:sp_sup_WKB}
\end{figure}

In the current work, we focus primarily on the propagation rather than the production of GWs.
Therefore, we have adapted the {\sc Pencil Code} to evolve an initial GW spectrum
in the absence of sources
with a spectral shape and amplitude based on those obtained in previous studies (see, e.g., refs.~\cite{He:2021bqm,RoperPol:2022iel}).
Hence, \Eq{eqn:GW_mod} is solved in one dimension
(i.e., in $k>0$) in order to improve the efficiency
of the code when studying the propagation of a GW background
along the history of the universe.
We take the initial spectrum for the time derivative of the strains,
defined in \Eq{eqn:Sh},
to be a smoothed double broken power law described by
\begin{equation}
S_{h'}(k, \eta_*) = S_{h'}^*
\frac{\biggl[1 + \Bigl(\frac{\kf}
{\kb}\Bigr)^{a - b}\biggr]
\Big(\frac{k}{k_*}\Big)^a}{\bigg[1 +
\Big(\frac{k}{\kb}\Big)^{(a - b)\alpha_1}\bigg]^{1\over \alpha_1}\bigg[1 +
\Big(\frac{k}{k_*}\Big)^{(b + c)\alpha_2}\bigg]^{1\over \alpha_2}
},
\label{sp_ini}
\end{equation}
where $S_{h'}^*$ and $k_*$ are approximately the initial amplitude and position of the peak, respectively, 
$\eta_* \approx \HH_*^{-1}$ corresponds to the time of GW production
during the RD era,
and $\alpha_1 = \alpha_2 = 2$ are fixed smoothness parameters.
We similarly initialize the spectrum of the strains $S_h (k, \eta_*) = S_{h'} (k, \eta_*)/k^2$.
We choose the slope in the IR range to be $k^2$, set by $a = 2$
(as is expected for causal sources 
of GWs like the ones produced
during a phase transition)
up to the break wave number $\kb = 1$, which corresponds to the horizon scale.
At intermediate wave numbers, $\kb \leq k \leq \kf$, the slope becomes
$b = 0$ as found, for example, for MHD turbulence in 
refs.~\cite{RoperPol:2019wvy,RoperPol:2022iel,Sharma:2022ysf},
followed by the power law $k^{-{11\over3}}$ in the UV range, set by $c = {11\over3}$.
This corresponds to the spectrum obtained for
Kolmogorov-like MHD turbulence \cite{RoperPol:2019wvy}.
Note, however, that this part of the spectrum corresponds to subhorizon
scales, which are described accurately by the WKB approximation
and, hence, the resulting spectral shape is not expected to be modified in this range.
On the other hand, around the horizon or at larger scales, depending
on the value of $\alpM^*$, the resulting GW spectral shape might be modified by the inclusion of
an additional IR branch $\xi^\WKB(k) \sim k^{-2}$ as
predicted in \Sec{GW_sp_WKB} by the WKB approximation.
Finally, the resulting spectrum is expected to be enhanced at all wave numbers
by a factor $e^{-2\ddd} (1 + \half \alpT)$ owing to the presence
of non-zero MG parameters $\alpM$ and $\alpT$.
\FFig{fig:sp_sup_WKB} shows the resulting GW spectrum estimated
using the WKB approximation for different values of $\alpM^*$.
Note, however, that the assumptions made by the WKB approximation break
down around the critical wave number at which we expect the relevant spectral modifications.

\begin{table}[t]
\centering
\begin{tabular}{c|c|r|r|r|c}
 & run & $\alpMz$ & $n$ & $\alpTz$ & $\EEGW/\EEGW^*$\\\hline
Choice 0   & T0A & $0$     & --     & $-0.5$ & $\ \,7.50\times10^{-1}$ \\
($\alpT$)  & T0B & $0$     & --     & $-0.2$ & $\ \,9.00\times10^{-1}$ \\
           & T0C & $0$     & --     & $0.2$  & $1.10\times10^0$ \\
           & T0D & $0$     & --     & $0.5$  & $1.25\times10^0$ \\\hline
Choice 0   & M0A & $-0.5$  & --     & 0      & $3.44\times10^7$ \\
($\alpM$)  & M0B & $-0.3$  & --     & 0      & $3.32\times10^4$ \\
           & M0C & $-0.1$  & --     & 0      & $3.21\times10^1$ \\
           & M0D & $-0.01$ & --     & 0      & $1.41\times10^0$ \\
           & M0E & $0.1$   & --     & 0      & $\ \,3.11\times10^{-2}$ \\
           & M0F & $0.3$   & --     & 0      & $\ \,3.03\times10^{-5}$  \\ \hline
Choice I   & M1A & $-0.5$  & $2$    & 0      & $1.28\times10^0$ \\
($\alpM$)  & M1B & $-0.3$  & $2$    & 0      & $1.16\times10^0$ \\
           & M1C & $-0.1$  & $2$    & 0      & $1.05\times10^0$ \\
           & M1D & $0.1$   & $0.4$  & 0      & $\ \,7.79\times10^{-1}$ \\
           & M1E & $0.3$   & $0.4$  & 0      & $\ \,4.73\times10^{-1}$ \\\hline
Choice II  & M2A & $-0.5$  & --     & 0      & $1.32\times10^0$ \\
($\alpM$)  & M2B & $-0.3$  & --     & 0      & $1.18\times10^0$ \\
           & M2C & $-0.1$  & --     & 0      & $1.06\times10^0$ \\
           & M2D & $0.1$   & --     & 0      & $\ \,9.46\times10^{-1}$ \\
           & M2E & $0.3$   & --     & 0      & $\ \,8.47\times10^{-1}$ \\\hline
Choice III & M3A & $-0.5$  & --     & 0      & $3.57\times10^8$ \\
($\alpM$)  & M3B & $-0.3$  & --     & 0      & $1.35\times10^5$ \\
           & M3C & $-0.1$  & --     & 0      & $5.12\times10^1$  \\
           & M3D & $-0.01$ & --     & 0      & $1.48\times10^0$  \\
           & M3E & $0.1$   & --     & 0      & $\ \,1.95\times10^{-2}$ \\
           & M3F & $0.3$   & --     & 0      & $\ \,7.48\times10^{-6}$ 
\end{tabular}
\caption{Parameters used for the numerical studies:
for all runs, $k_1 = 10^{-3}$ is the smallest wave number and
$N = 46\,000$ is the number of grid points in one dimension.
$\EEGW/\EEGW^*$ indicates the 
present-day values of the energy density due to MG with respect to the corresponding GR values.
For each of the runs in series M0 and M3, we have performed an additional run with $k_1 =
10^{-7}$ to study the spectral modifications at even smaller $k$.}
\label{tab:run_params}
\end{table}

Table~\ref{tab:run_params} summarizes the parameters of the runs for the numerical studies.
The values of $\alpTz$ are chosen to be much larger than the current constraints
in order to later show the relative insignificance of
$\alpTz$ even when it takes large values.
The choices of $\alpMz$ are made in line with the limits discussed in \Sec{sec:param}.
For the choice II of $\alpM$ parameterization, we take $n = 2$ for
negative values of $\alpMz$ and $n = 0.4$ for positive values to
ensure that the stability conditions of \Eq{stab_cond} hold.

For series T0 (T0A--T0D),
we evolve the solution entirely with increasing time steps,
such that $\eta_{\rm next} = \eta_{\rm current}(1 + \del n_\incr)$ 
with $\del n_\incr = 0.01$, leading to equidistant points in logarithmic time spacing.
For series M0 (M0A--M0F), M1 (M1A--M1D), M2 (M2A--M2D), and M3 (M3A--M3F), 
we keep the nonuniform time scheme during RD and MD but switch to linear
time steps during $\Lam$D such that $\eta_{\rm next} = \eta_{\rm current}
+ \del\eta$ with $\del\eta \, \HH_* = 5\times10^9$.
We choose such time evolution to
improve the accuracy of late-time results for the cases of time-dependent $\alpM$,
especially the modifications that they present in the IR limit.
We show in \App{sec:numerical_accuracy} that
decreasing the time step below $\delta \eta \, \HH_* = 5\times 10^9$
does not affect the IR range of the spectra, which
indicates that the observed modifications are not due to numerical inaccuracy.
Since series T0 does not exhibit $k$-dependent modifications, it
does not require the additional
computational effort.
The choice of time schemes and their numerical accuracy is further discussed
in \App{sec:numerical_accuracy}.

In the simulations, we consider the initial time to be the EWPT
with a temperature scale $T_* \sim 100\,$GeV and the number of relativistic and 
adiabatic degrees of freedom are $\gS \approx g_* \sim 100$,
which yields the values of $a_*$ and $\HH_*$ in \Eqs{as_a0}{HHs}.
With these values, we can then solve \Eq{eqn:GW_mod} in units of
the normalized time $\eta \HH_*$, by mapping the parameterizations in \Eq{eqn:alpM_param} $\alpM(\eta) \map \alpM(\eta/\eta_*)$ and using the results from Friedmann equations for
$\HH$ and $a''/a$ (which are already normalized since $a'$ is
computed as the derivative with respect to $\eta \HH_*$),
given in \Eq{eqn:a_derivatives_omega}.
The effects on $S_{h'} (k, \eta)$ of the specific choice of the time at which the
GWs are generated only appear via the relative magnitude of the terms $a''/a$
and $\alpM'$ that involve time 
derivatives
compared to $k$.
These terms have been parameterized in \Eqs{rel_app}{rel_alpMp},
respectively, and their magnitude has been discussed.
We have shown that the term $a''/a$ can only induce
modifications to the solution at scales several orders of magnitude
above the horizon scale, while $\alpM'$ is only of the order of
$\alpM$ for choice III (see \Sec{sec:param}) at very late times within
the RD era.
For other choices of $\alpM (\eta)$, either $\alpM' = 0$
(choice 0) or $\alpM$ itself is orders of magnitude below its present-time value $\alpMz$ (choices I and II).
We find that the WKB approximation
is expected to be valid around $k \gg \klimaM \sim \sqrt{|\alpM^*|/2}$ and, on the other hand,
it predicts an enhancement $k^{-2}$ in the IR regime $k \lesssim k_\crit \leq \klimaM$, which is not on the range
of validity of the WKB estimate.
Using the results of the numerical simulations, we find in \Sec{ssec:energy_spectrum} an
IR enhancement $\xi (k) \propto k^{-\beta_0}$ with $\beta_0 \sim 2$ and $\beta_0 \gtrsim 0$ for negative and positive values of $\alpMz$, respectively, 
that does
not
in general follow the spectral shape predicted by WKB and that
can become shallower at smaller $k$.
The position of this spectral change is a fixed fraction of
the Hubble horizon at the
time of generation, determined by the value of $\alpM^*$, that
does not depend on the specific value of $\eta_*$.
Finally, the time of generation determines the range of frequencies
where we observe the signal as well as its amplitude via \Eq{eqn:OmGW_Sh}.

The amplitude $S_{h'}^*$ and the position of the peak $k_*$ can be chosen to represent a specific
model.
For example, the values $S_{h'}^* = 5 \times 10^{-10}$ and $k_* = 10$
are used in our runs (see table~\ref{tab:run_params}) and they would
produce a normalized initial total energy density\footnote{
The turbulent and GW energy densities are normalized to the 
radiation energy density, such that $\EEGW$ during RD can be
computed as
\begin{equation}
     \EEGW (\eta) = \frac{1}{6} \int_0^\infty S_{h'} (k, \eta) \dd k. \nonumber
\end{equation}
For the spectral shape $S_{h'} (k)$ given in \Eq{sp_ini} with $k_* = 10$, one finds $\EEGW \simeq 2 S_{h'}^*$ (see the associated online material \cite{GH}).
}
$\EEGW^*\simeq 10^{-9}$,
which corresponds to a vortically turbulent source with an energy density of roughly
$q \eee_\turb^* \simeq 1.74 \times 10^{-4}$.
This is related via
\begin{equation}
\EEGW^* = (q\eee_\turb^*/k_{\rm s})^2,
\label{eqn:EEGW_efficiency}
\end{equation}
where $q$ is an empirically determined coefficient for a 
specific type of turbulence source $\eee_\turb^*$, and is found
to be of the order of unity or larger (up to $\sim$\,5), depending on the production
mechanism of the source 
\cite{RoperPol:2019wvy,Brandenburg:2021bvg,RoperPol:2021xnd},
and $k_{\rm s}$ is the characteristic scale of the turbulence
sourcing, which is related to $k_*$ in \Eq{sp_ini} as $k_* 
\simeq 1.143 \times 1.6 \, k_{\rm s} \simeq 1.83\, k_{\rm s}$,
where the factor 1.143 relates the position
of the maximum of $\OmGW (k) \propto k S_{h'} (k)$ and 
$k_*$ (see the associated online material \cite{GH})
and $1.6$ the relation between $k_{\rm s}$ and the maximum \cite{RoperPol:2022iel}.

\subsection{Time evolution}

For constant $\alpT \neq 0$ and $\alpM = 0$, the GW energy density
stays constant in time and its magnitude is
modified by the specific value of $\alpT$, as it can be predicted from the analytical solution to \Eq{eqn:GW_mod}
in the absence of sources during RD, given in \Eq{xi_mod_general} under the
WKB approximation.
We find excellent agreement between the WKB estimate and the
numerical solution of the GW energy density.
Note that this corresponds to a boost of energy for
$\alpT > 0$ and a depletion for $\alpT < 0$, given by $1 + \half \alpT$.
Hence, the relative changes upon the GR solutions are of order
$\alpT \sim \ooo(10^{-1})$
(see \Tab{tab:run_params}),
which would become much smaller if we restrict $\alpT$ to the present-time
constraint at the LIGO--Virgo frequency band $\alpTz\lesssim\ooo(10^{-15})$.

\begin{figure}[t]
\centering
\includegraphics[width=0.490\textwidth]{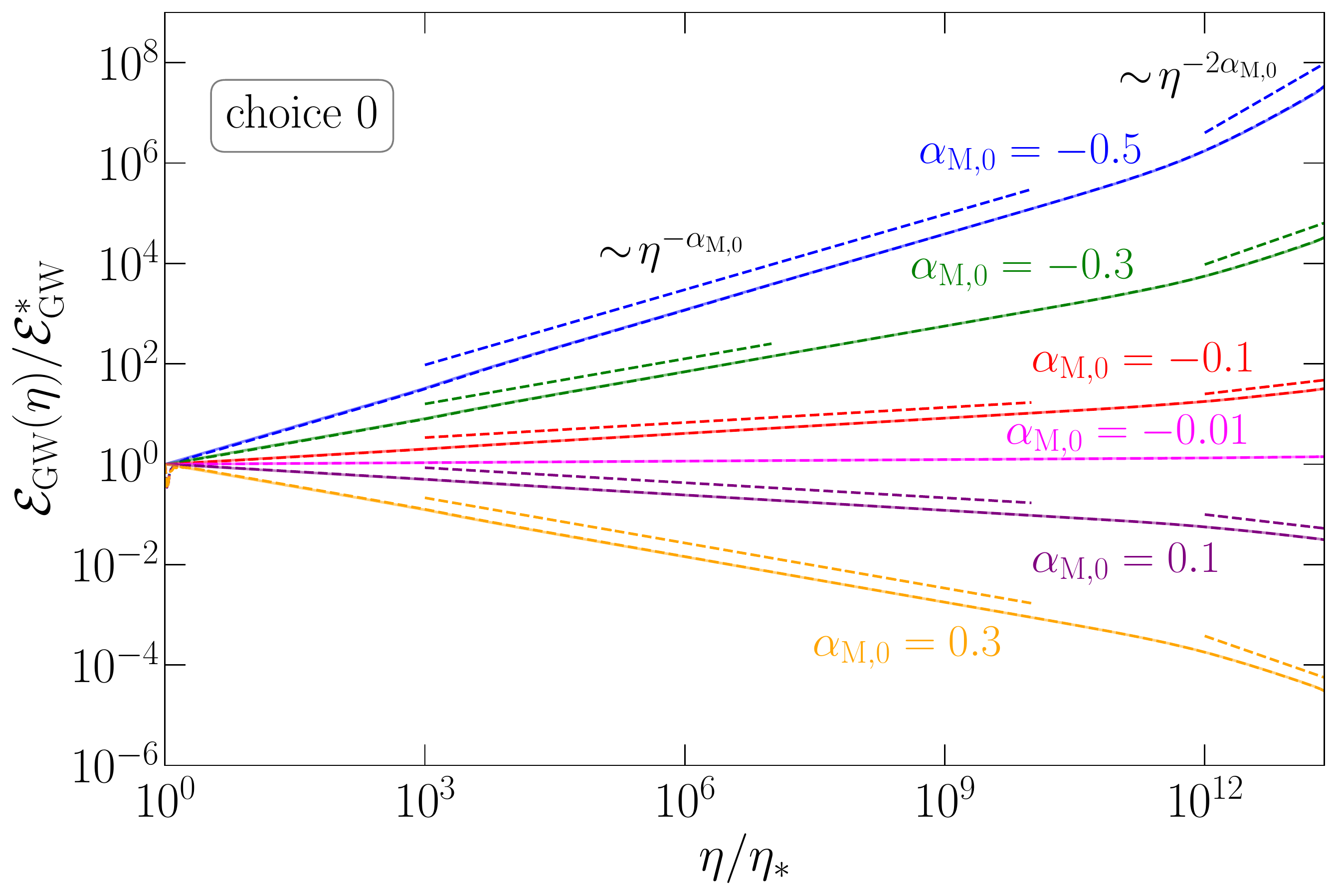}
\includegraphics[width=0.502\textwidth]{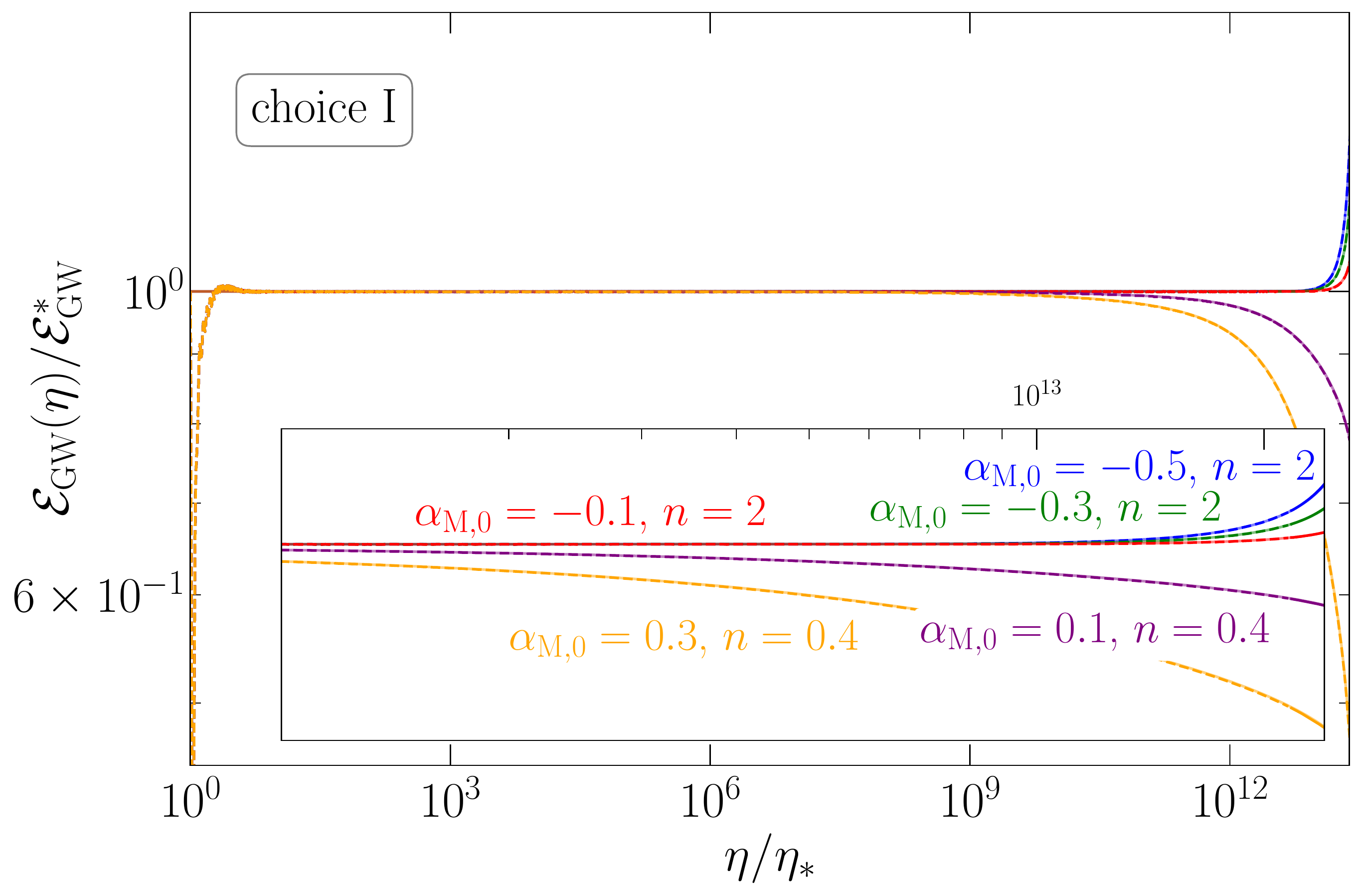}
\includegraphics[width=0.502\textwidth]{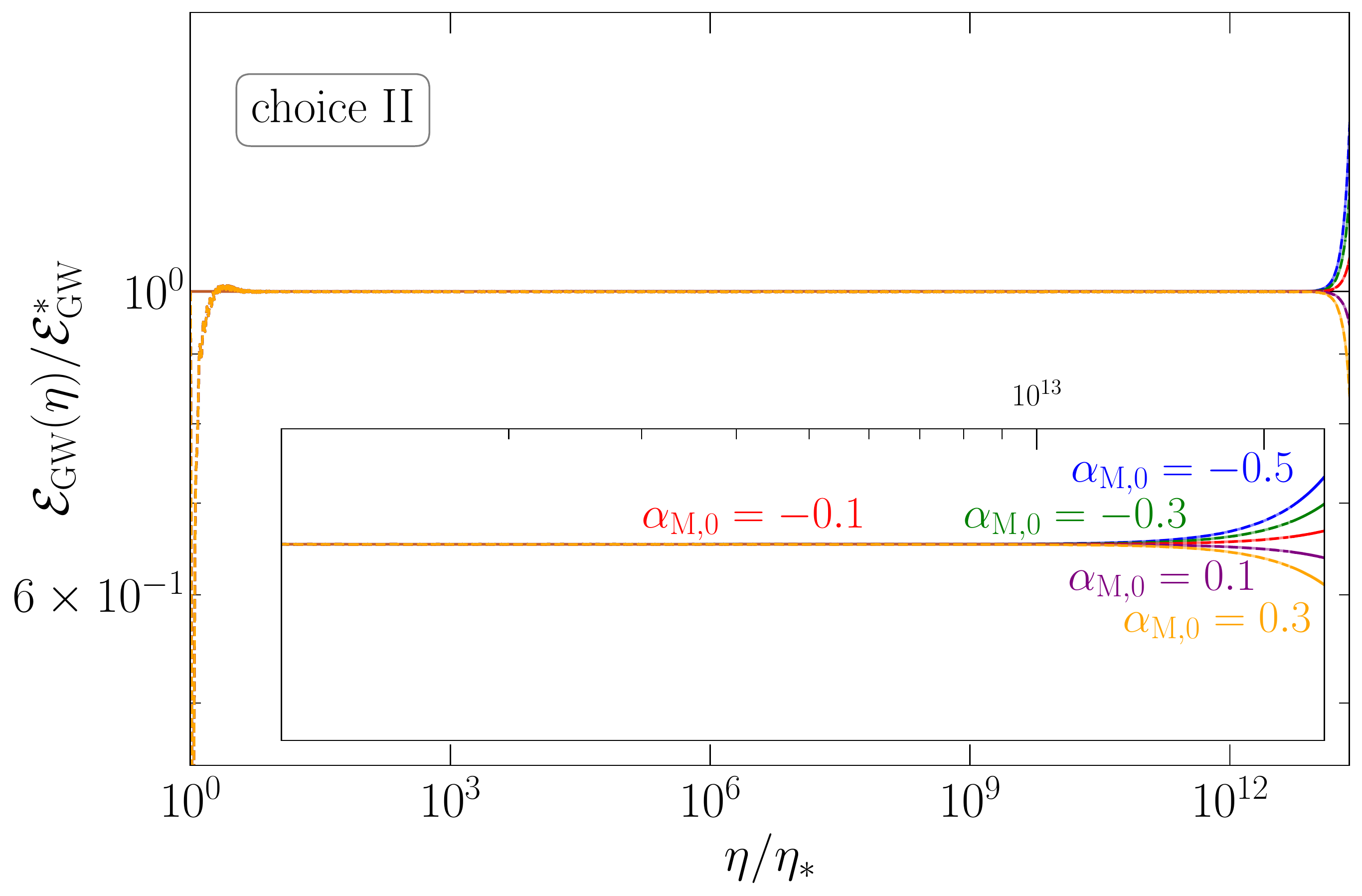}
\includegraphics[width=0.490\textwidth]{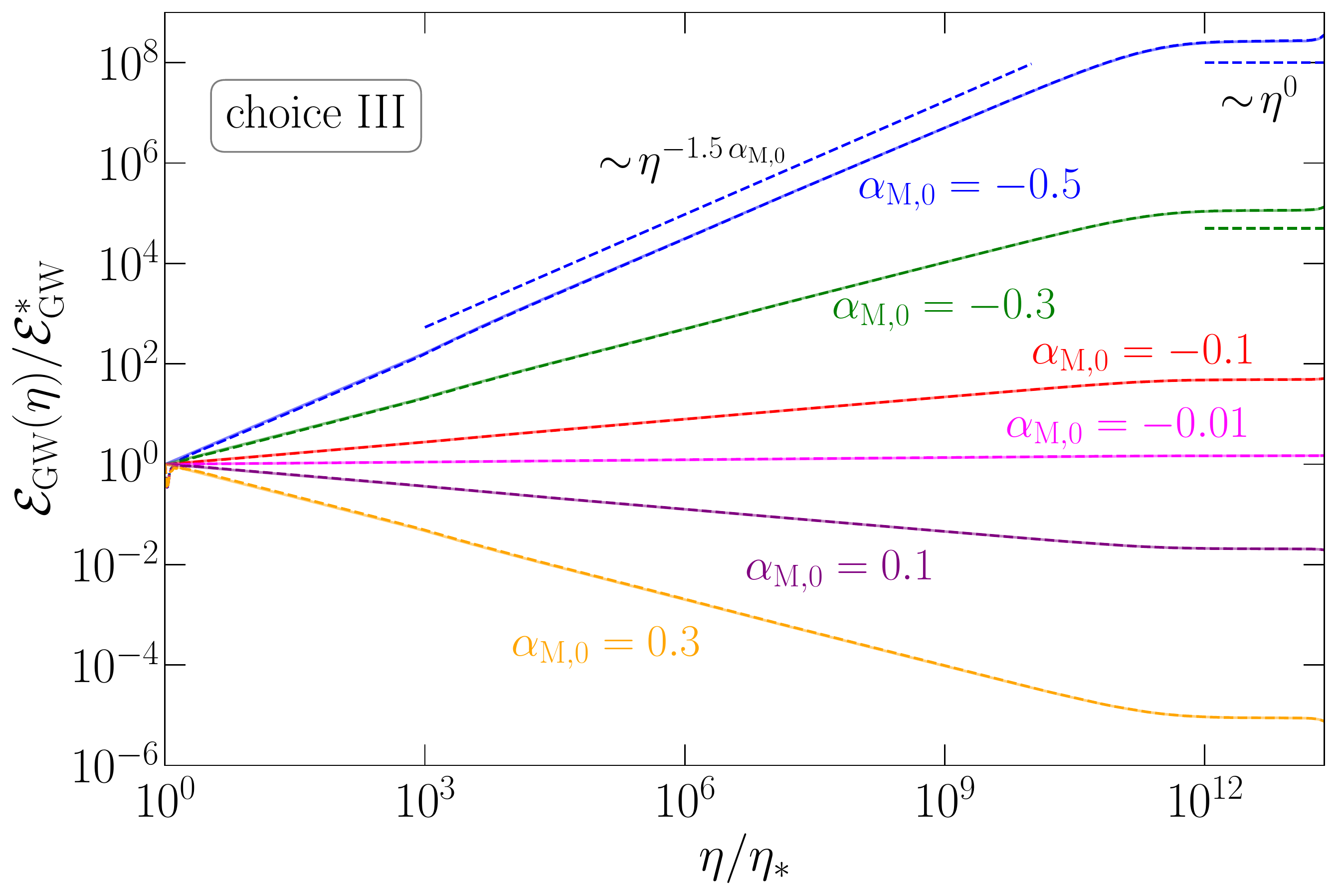}
\caption{Time evolution of the normalized total energy density $\EEGW/\EEGW^*$
for choices 0 to III and $\alpMz \in \{-0.5, -0.3, -0.1, -0.01, 0.1, 0.3\}$.
In all panels, the WKB estimates are shown as solid lines and
numerical solutions as dashed lines, being almost
indistinguishable.
For choices I and II, the inset corresponds to the $\Lambda$D era,
when there are relevant modifications to the GR solution $\EEGW^*$.
All runs are initialized at the EWPT such that the present time is $\eta_0/\eta_*\simeq 2.4\times10^{13}$.}
\label{fig:ts_all}
\end{figure}

\FFig{fig:ts_all} presents the time evolution of the total GW 
energy density $\EEGW(\eta)$ of the runs in series M0 (upper left), M1 (upper right), M2 (lower left), and M3 (lower right).
In these runs, $\alpM$ follows each of the parameterizations given
in \Eq{eqn:alpM_param} and $\alpT = 0$.
Regardless of the specific parameterization,
the numerical solutions and the WKB 
approximations agree on the total GW energy
density time dependence, enhanced or depleted by a factor
$e^{-2\ddd}$ for negative or positive values of $\alpMz$, respectively.
\begin{figure}[t]
\centering
\includegraphics[width=.490\textwidth]{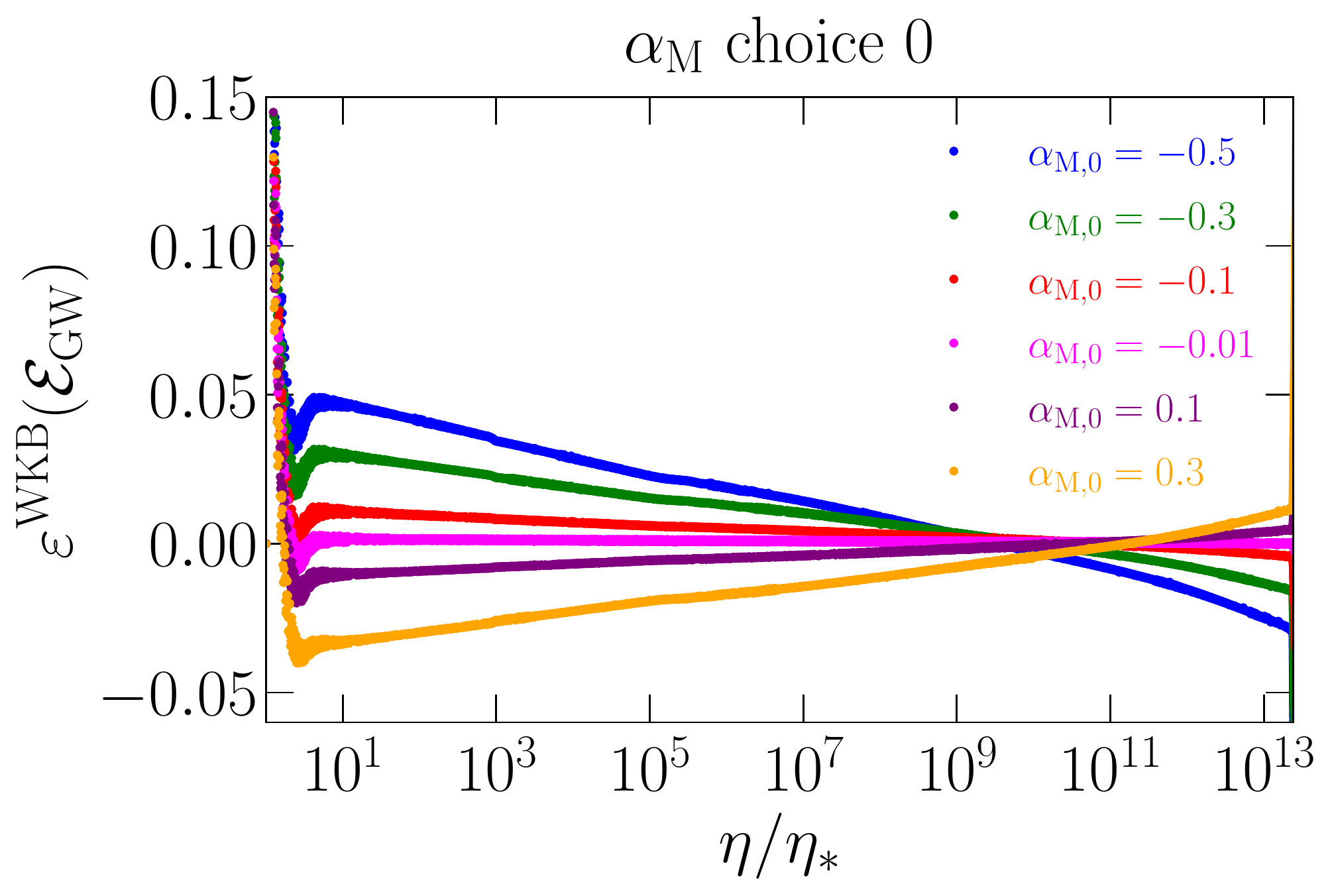}
\includegraphics[width=.502\textwidth]{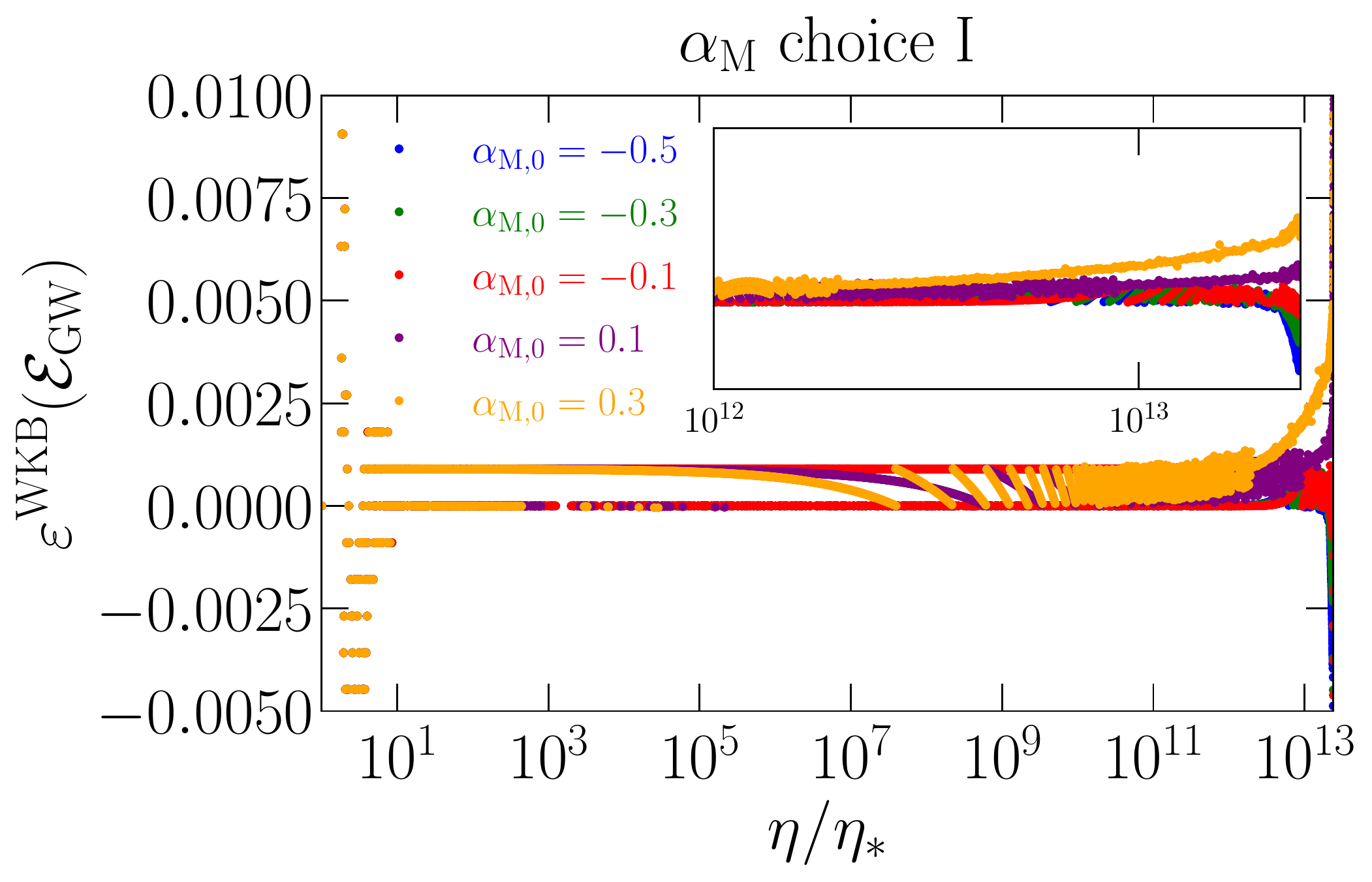}
\includegraphics[width=.490\textwidth]{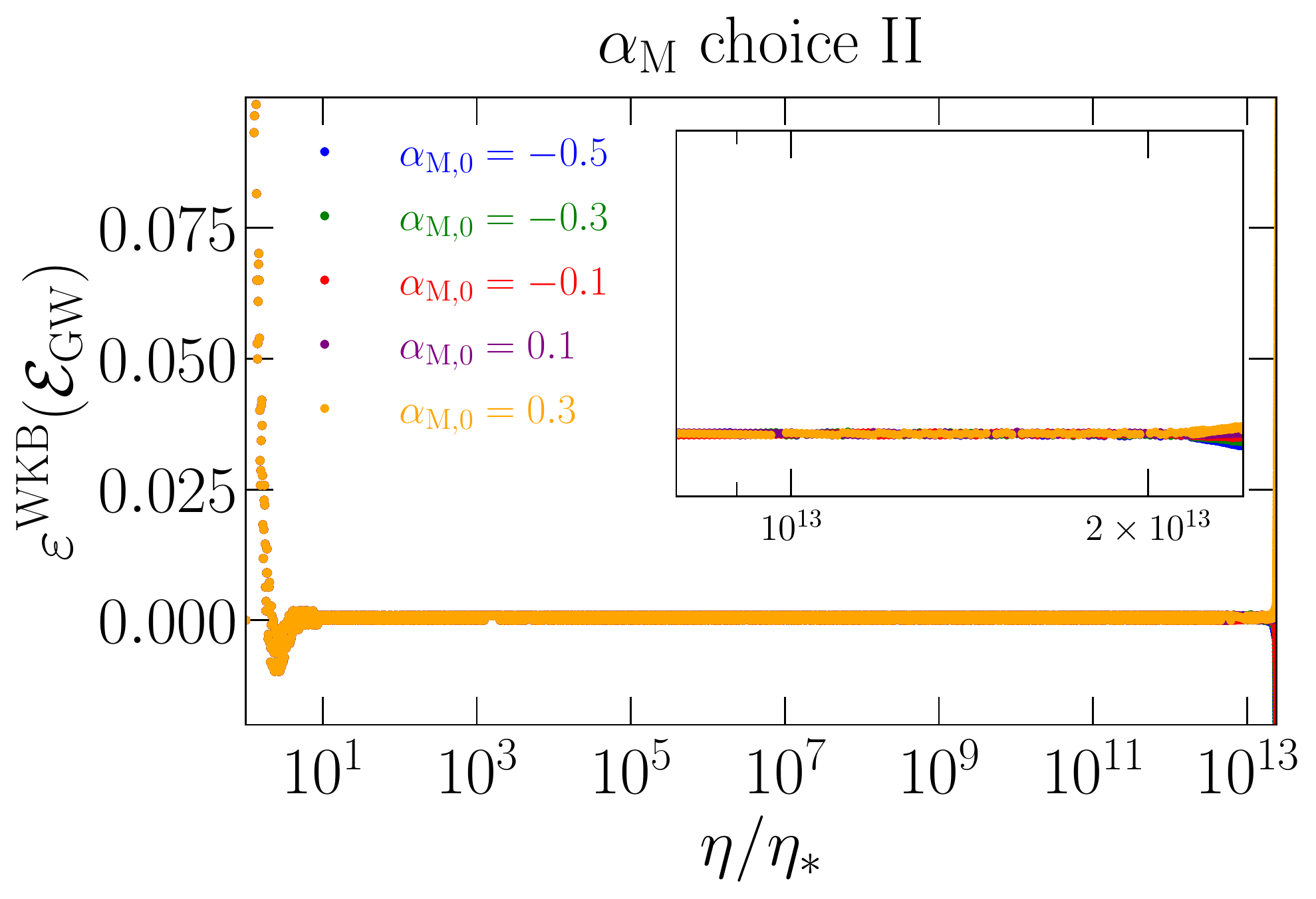}
\includegraphics[width=.502\textwidth]{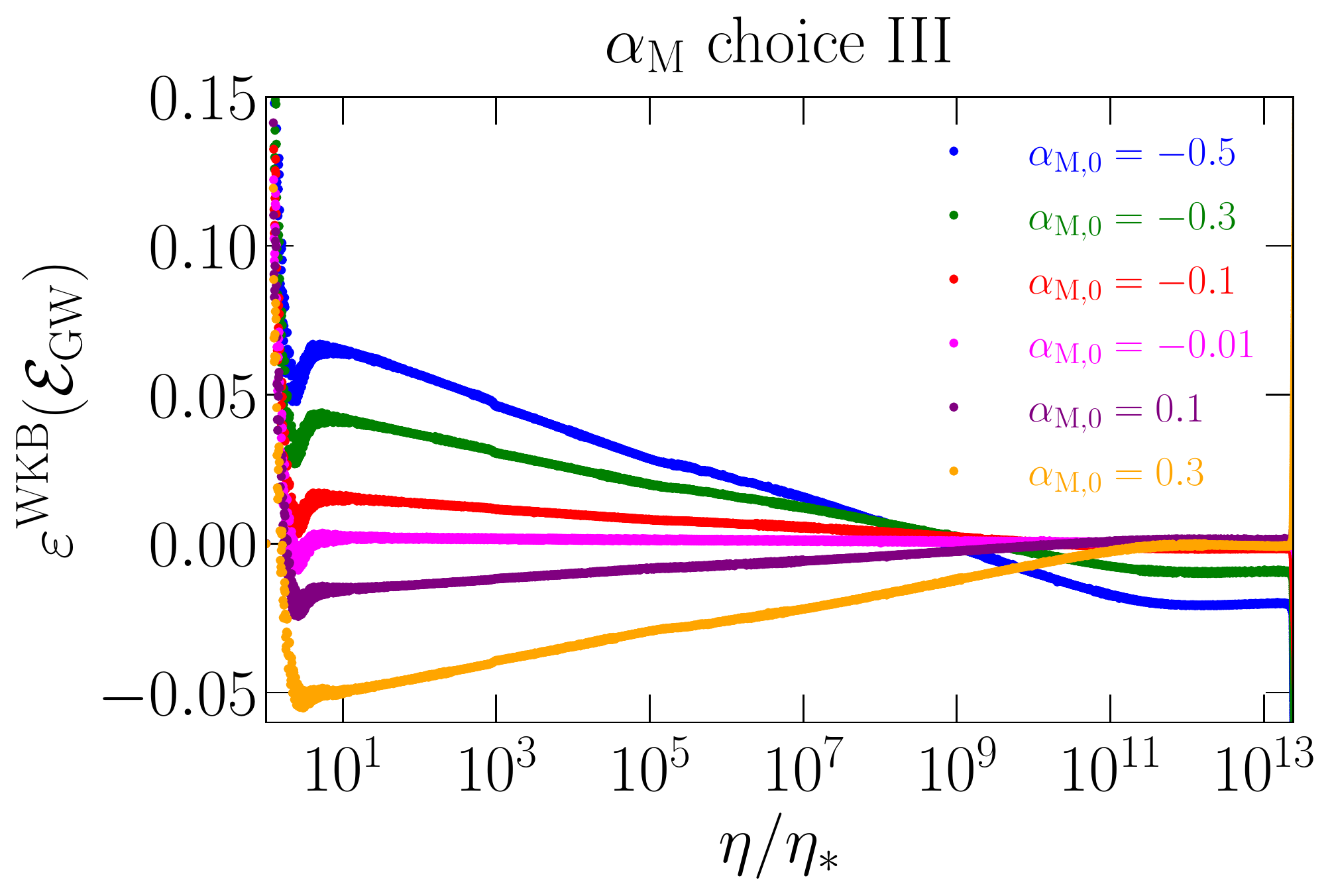}
\caption{Relative error on the time series in \Fig{fig:ts_all}
of the WKB approximation compared to the numerical simulations,
$\varepsilon^\WKB ({\cal E}_\GW)$.}
\label{fig:ts_error_WKB}
\end{figure}

Choices 0 and III yield similar results, with a time evolution
$\EEGW (\eta) \sim \eta^{-\alpMz}$ and $\EEGW (\eta) \sim \eta^{-1.5 \, \alpMz}$ during RD era, and $\EEGW \sim \eta^{-2 \alpMz}$ and
$\EEGW (\eta) \sim \eta^0$ during MD, respectively.
This agrees with the expectation since choice III mostly differs from choice 0 during MD.
During RD, both parameterizations are almost the same and just differ
by a factor $1/(1 - \Om_{\mat, 0}) \simeq 1.46$, leading to the different
scalings.
In other words, the evolution of $\EEGW$,
proportional to $a^{-\alpM}$ when $\alpM$ is constant (e.g., for choice 0, and for choice III during RD era),
is determined by the cosmic expansion itself.
Note that choices 0 and III induce an enhancement or suppression in the total energy density
that heavily depends on the values of $\alpMz$,
i.e., $\EEGW/\EEGW^*$ can range from $\ooo(10^{-5})$ (choice 0) and $\ooo(10^{-6})$ (choice III) for
$\alpMz = 0.3$ 
to $\ooo(10^7)$ (choice 0) and $\ooo(10^8)$ (choice III) for $\alpMz = -0.5$.
In general, choice III leads to a larger modification due to larger
values of $\alpM$ during RD than choice 0.
The potential implications of such a large GW energy density
enhancement are discussed in \Sec{sec:obs_implications}.

For both choices I and II (see the upper right and lower left panels of \Fig{fig:ts_all}),
the modified GW solutions remain close to their GR counterparts for most of the time,
and rapidly depart from GR as $\eta$ enters $\Lam$D and approaches the present day.
This is expected since $\alpM$ is
proportional to the scale factor and the dark energy density
for choices I and II, respectively, and hence, the values of $\alpM$ are negligibly small for most of the cosmic
history until $\Lam$D era.
For this reason, for the same values of $\alpMz$,
the final values of $\EEGW$ in both of these cases are significantly lower than those in choices 0 and III,
where the modifications are accumulated from RD onward.

In \Fig{fig:ts_all}, the differences between the WKB and the numerical solutions are indistinguishable.
We quantify in \Fig{fig:ts_error_WKB} the relative error between the two,
defined to be $\varepsilon^\WKB(\EEGW)\equiv[\EEGW^\WKB (\eta) - \EEGW (\eta)]/\EEGW(\eta)$.
In all panels, there exists a brief but relatively large error region,
amounting to $\lesssim15\%$,
around the initial time $\eta_*$.
This is due to the sinusoidal oscillations of each $k$ mode that are present in GR and MG alike [see \Eq{simp_WKB}].
It is a consequence of setting an already saturated spectra
$S_{h'} (k)$ as the initial condition at time $\eta_*$ and hence, it is not due to the WKB estimate.
After the initial oscillations settle down,
$\varepsilon^\WKB(\EEGW)$ for choices 0 and III decrease over the 
majority of time in RD era.
Although they both increase somewhat later on,
the maximum error at the final time is only $\varepsilon^\WKB(\EEGW)\sim3\%$.
For choices I and II, due to the negligible impact of $\alpM$ during RD,
the GW solutions settle down to the same magnitude after the initial oscillations.
Therefore, the relative errors during RD remain close to zero.
During $\Lam$D, the errors grow as the effects of $\alpM$ become more significant.
But this is also limited, since at most
$\varepsilon^\WKB(\EEGW)\sim1\%$ is found at the present day,
lower than in choices 0 and III.

\subsection{Energy spectrum}
\label{ssec:energy_spectrum}

\begin{figure}[t]
\centering
\includegraphics[width=.496\textwidth]{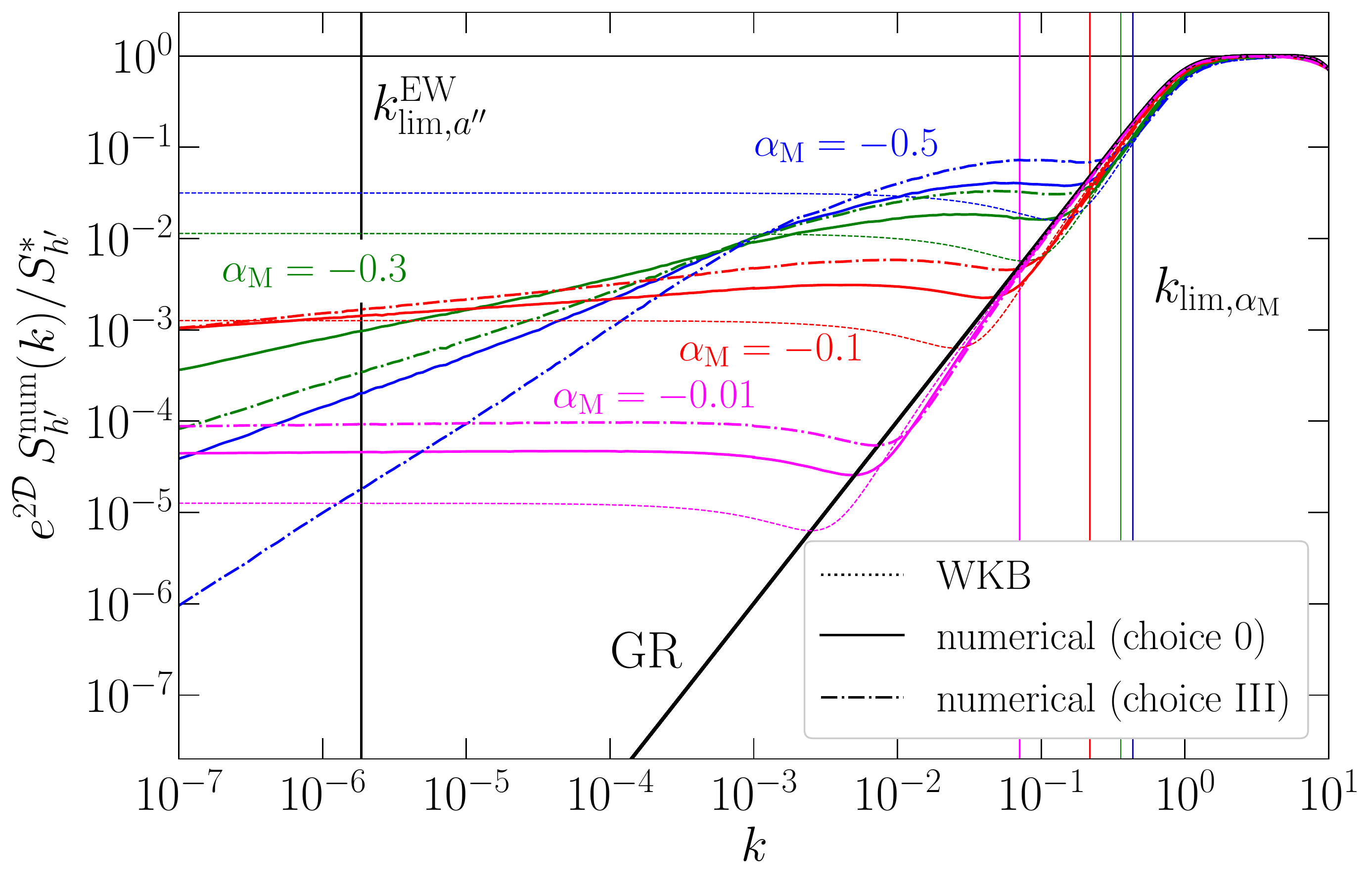}
\includegraphics[width=.496\textwidth]{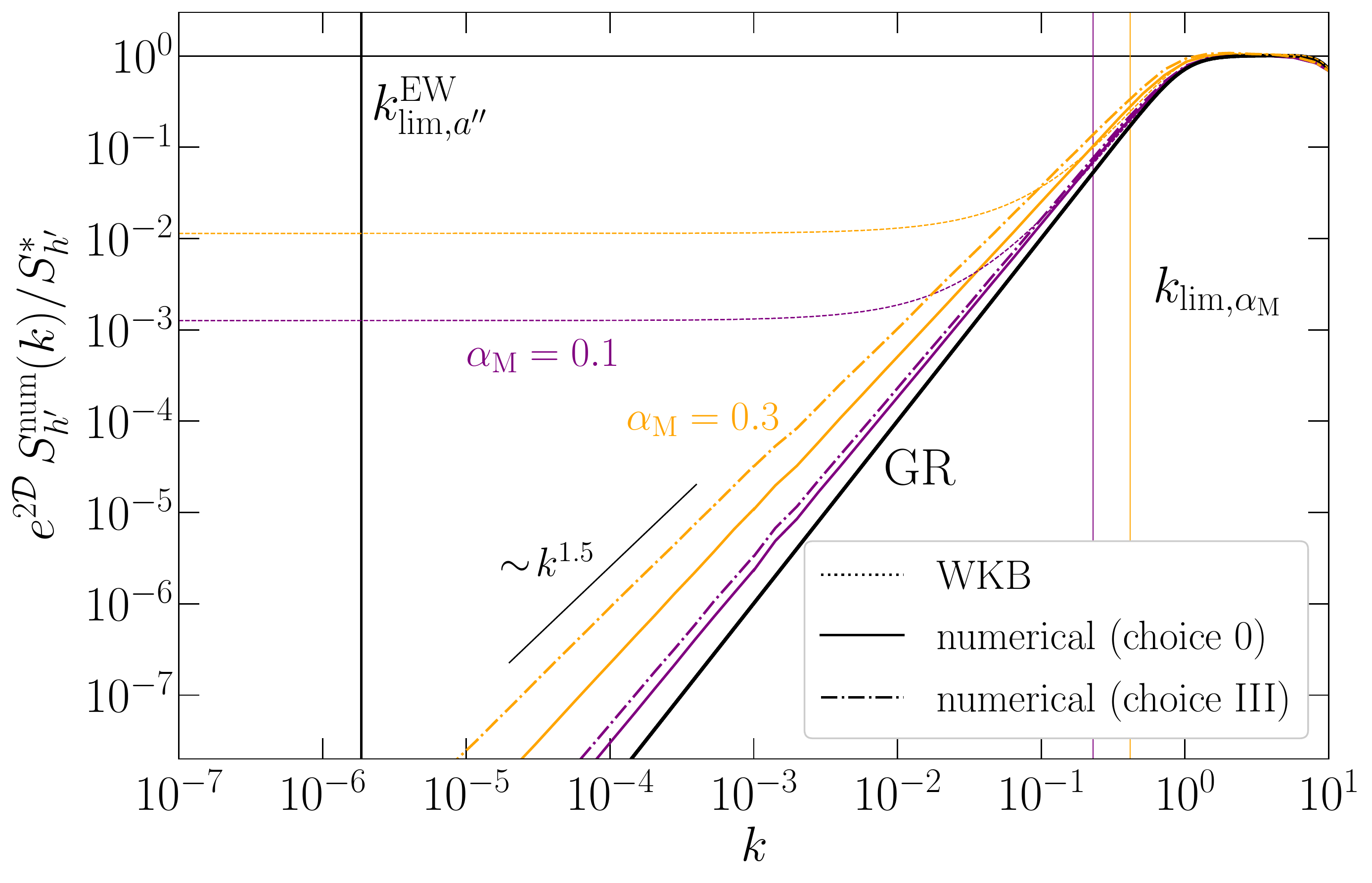}
\caption{Saturated final energy spectra compensated by the change on the
total GW energy density
$e^{2\ddd}S_{h'}^{\rm num} (k)/S_{h'}^*$ for the numerical runs with the choices 0 and III of 
$\alpM$ parameterization.
Left and right panels represent negative and positive values of $\alpMz$, respectively.
The GR solution and WKB approximation are shown in black 
solid and color dotted lines, respectively.
Numerical solutions are shown in solid (choice 0) and dash-dotted (choice III) lines.
Thin vertical lines indicate the corresponding limits to the WKB
approximation $\klimap^\EW$ and $\klimaM$.
The horizontal lines show the values at the peak $S_{h'}^*$.}
\label{fig:sp_all}
\end{figure}

In \Fig{fig:sp_all}, we show
the final energy spectrum averaged over oscillations and compensated by the factor $e^{-2\ddd}$
to study the changes on the spectral shape.
The saturated spectra are shown for the choices 0 and III of $\alpM$
parameterization,
since the rest of runs ($\alpT$ choice 0, $\alpM$ choices I and II) 
exhibit the same spectral shape as in GR.
To study the spectral changes, we again define $\xi (k) = e^{2\ddd} S_{h'} (k)/S_{h'}^\GR (k)$, in analogy to \Eq{OmGW_mod_general}.
As a reminder of the potential limitations of the WKB approximation,
we mark with thin vertical lines the values of $\klimap^\EW$ and $\klimaM$,
and note that the latter occurs at wave numbers larger than the IR regime characterized by $k_\crit = \half |\alpM^*|$.

\begin{figure}[t]
\centering
\includegraphics[width=.501\textwidth]{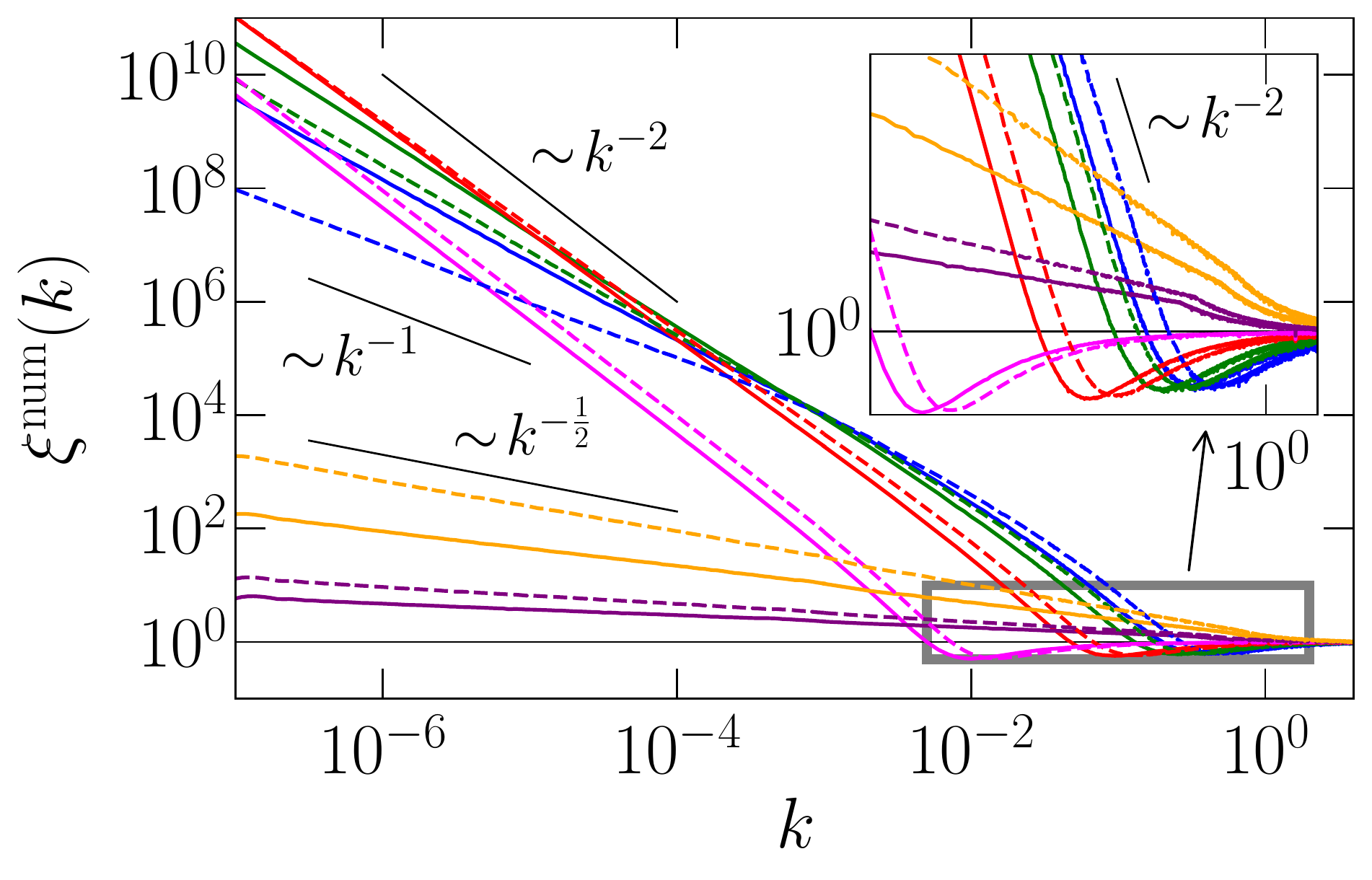}
\includegraphics[width=.491\textwidth]{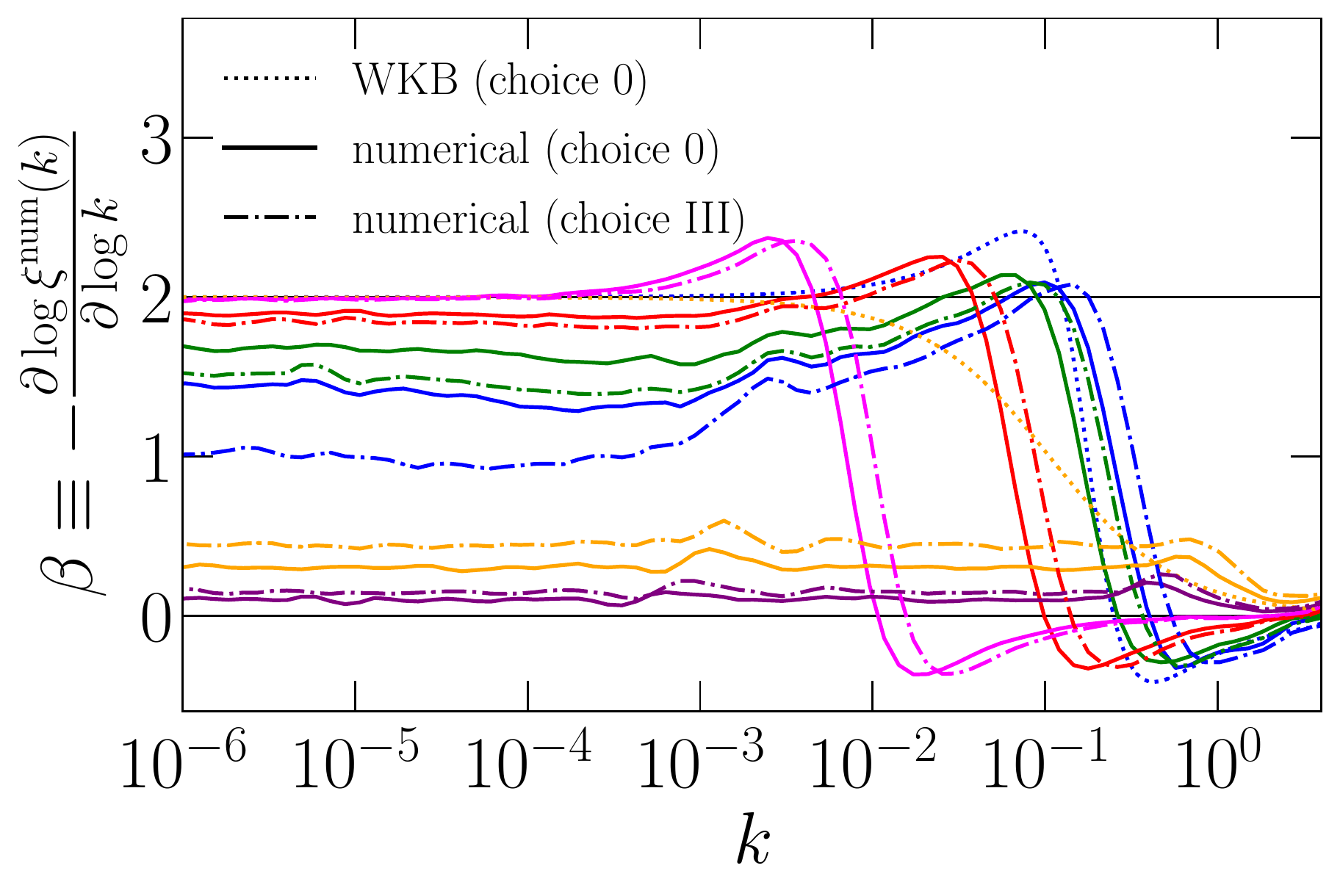}
\caption{Left and right panels respectively show 
$\xi^\num (k)$,
the saturated spectra normalized with the GR solution, and $\beta$, its spectral slope as a function of wave number.
We show, for comparison, the slopes $\beta$ predicted by the WKB estimate for
$\alpMz = -0.5$ and $0.3$.
The horizontal thin black lines indicate the GR solution in the left panel and
the slopes $0$ and $2$ in the right panel, for reference.
Different values of $\alpMz$ and line colors are consistent with the ones used in 
\Figss{fig:ts_all}{fig:sp_all}.}
\label{fig:sp_norm}
\end{figure}

Comparing the numerical results to the expected spectra obtained
using the WKB approximation (see \Fig{fig:sp_sup_WKB}),
we note two main differences.
In the first place, we find an IR enhancement
$\xi (k)\sim\!k^{-\beta_0}$ with $\beta_0 
\in (1, 2)$, where $\beta_0 = 2$ corresponds to the WKB approximation,
for negative values of $\alpMz$,
whereas for positive values there are more moderate modifications with $\beta_0 \lesssim 0.5$.
Moreover, even for negative $\alpMz$, the modifications to the spectral shape are
different than those predicted by the WKB estimate.
These modifications can be seen more clearly in \Fig{fig:sp_norm},
where the left panel shows the numerically obtained $\xi (k)$
and the right panel directly presents the spectral slopes of the compensated spectra
$\beta \equiv - \partial \log \xi (k)/\partial \log k$, measured at each wave number.

The slope $\beta$ of the WKB estimate can be computed from \Eq{xi_mod_general} as
\begin{equation}
    \beta = \frac{2 \alpM^*\bigl(\alpM^* + 2 k\bigr)}
    {8k^2 + \alpha^{*2}_{\rm M} + 4 k \alpM^*},
    \label{slope_beta}
\end{equation}
which
has an initial departure $\beta < 0$ for $\alpM^* < 0$ as $k$ decreases
until it reaches a peak $\beta_{\rm min} = 1 - \sqrt{2} \simeq -0.41$ at $k_{\beta_{\rm min}} =
\fourth \alpM^* \sqrt{2}/(1 - \sqrt{2}) \simeq 0.85 \, |\alpM^*|$.
It then grows again to zero at
$k_\crit$ and continues to increase up to $\beta_{\rm max} = 1 + \sqrt{2} \simeq 2.41$
at $k_{\beta_{\rm max}} = -\fourth \alpM^* \sqrt{2}/(1 + \sqrt{2}) \simeq 0.15 \, |\alpM^*|$,
crossing the aforementioned IR enhanced slope of $-2$ at the wave number
$k_{\beta = 2} = \fourth |\alpM^*|$ (for details, see the associated online material \cite{GH}).
Finally, the slope becomes slightly shallower and asymptotically tends to $-2$ as $k \rightarrow 0$, as can be seen by taking the limit in \Eq{slope_beta}.
For positive values of $\alpM^*$, the slope $\beta$ monotonically increases from $0$ to $2$ with decreasing $k$
so it does not present any local minima or maxima.
The slope $\beta$ for $\alpM^* = -0.5$ and $0.3$ under the WKB
estimate is shown in the right panel of \Fig{fig:sp_norm} for comparison with the numerical
results.

At wave numbers close to where the IR regime starts, $k\lesssim k_\crit$,
the $k^{-2}$ enhancement is reproduced by the numerical results
for negative $\alpMz$,
flattening that part of the original GR spectrum to $k^0$ (see \Fig{fig:sp_all}).
But as $k$ decreases,
the numerical spectra exhibit slopes steeper than $k^0$,
closer to the original GR $k^2$ spectrum,
that are not predicted by the WKB approximation, which maintains the
$k^0$ slope throughout all scales $k\lesssim k_\crit$.
The departure in the IR regime occurs at different scales depending on the values of $\alpMz$.
For $\alpMz = -0.01$, this is the least obvious,
where the enhanced spectrum still keeps a roughly $k^0$ shape even at $k\ll k_\crit$.
For $\alpMz = -0.5$, however,
the difference becomes visible when $k$ is still relatively close to the start of the IR regime at $k\sim k_\crit$.
We characterize the differences of the numerical results with those
from the WKB estimate in \Tab{tab:kcrit_beta} by computing the critical $k$ at which
$\beta = 2$, $k_{\beta=2}^\crit$.
Under WKB this only occurs at $\fourth |\alpM^*|$ and in the limit $k \rightarrow 0$ while in the
simulations it occurs at a similar $k$ and then again at a
larger but finite scale $k_{\beta=2}^{\crit, 2}$, when the spectra becomes steeper (with respect to the
flat spectrum induced by $\xi^{\rm WKB} (k) \sim k^{-2}$).
We also give the final slope $\beta_{0}$, such that $\xi (k)
\sim k^{-\beta_0}$ as $k \rightarrow 0$.

On the other hand, for positive $\alpMz$, the IR enhancement predicted by
the WKB estimate is not observed in the numerical results.
In this case, the modified spectra present small increases around
$k\gtrsim1$ and slightly shallower
slopes than their GR counterparts at all superhorizon scales,
characterized by $\beta_0 \gtrsim 0$.

\begin{figure}[t]
\centering
\includegraphics[width=0.496\textwidth]{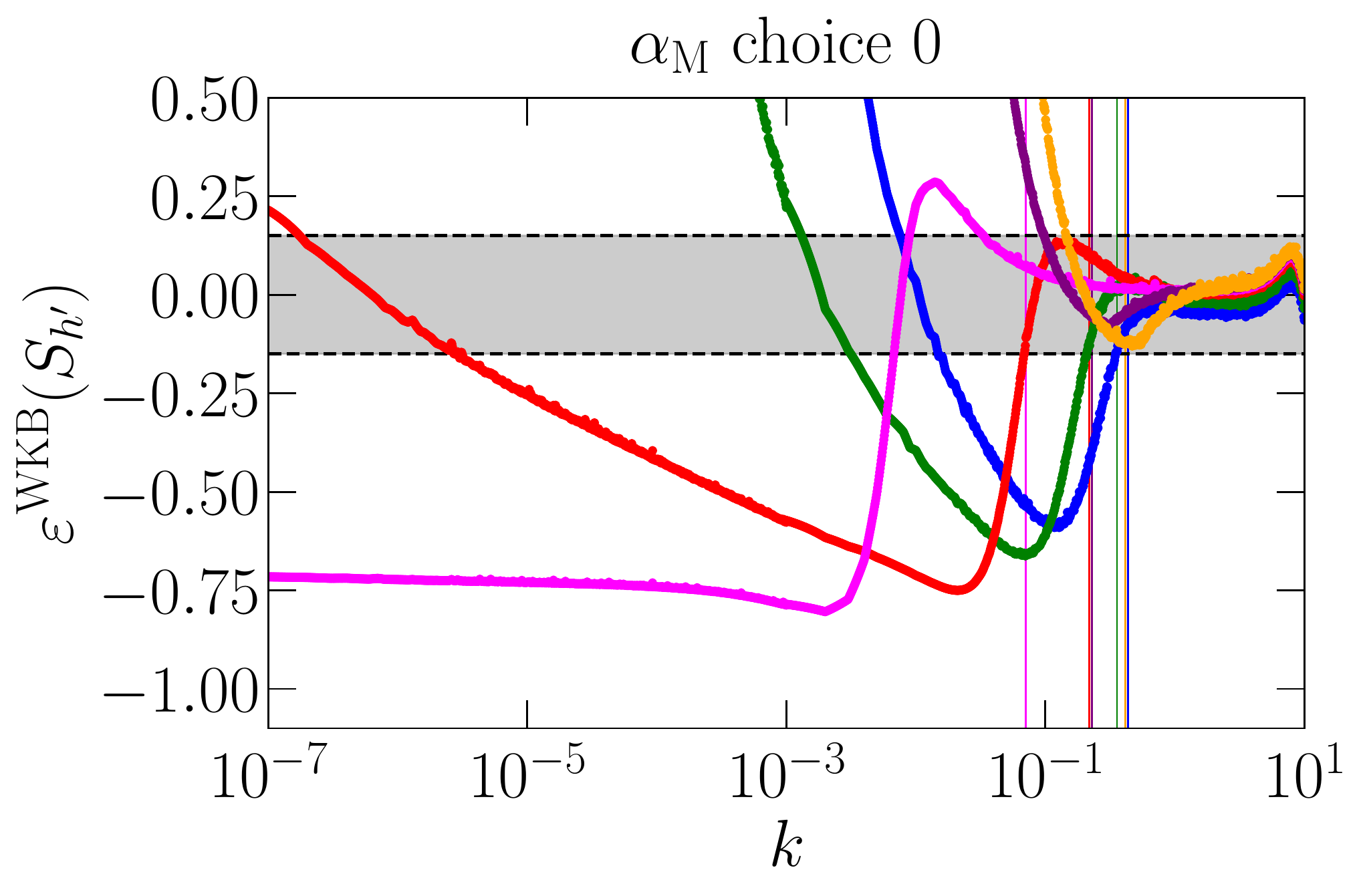}
\includegraphics[width=0.496\textwidth]{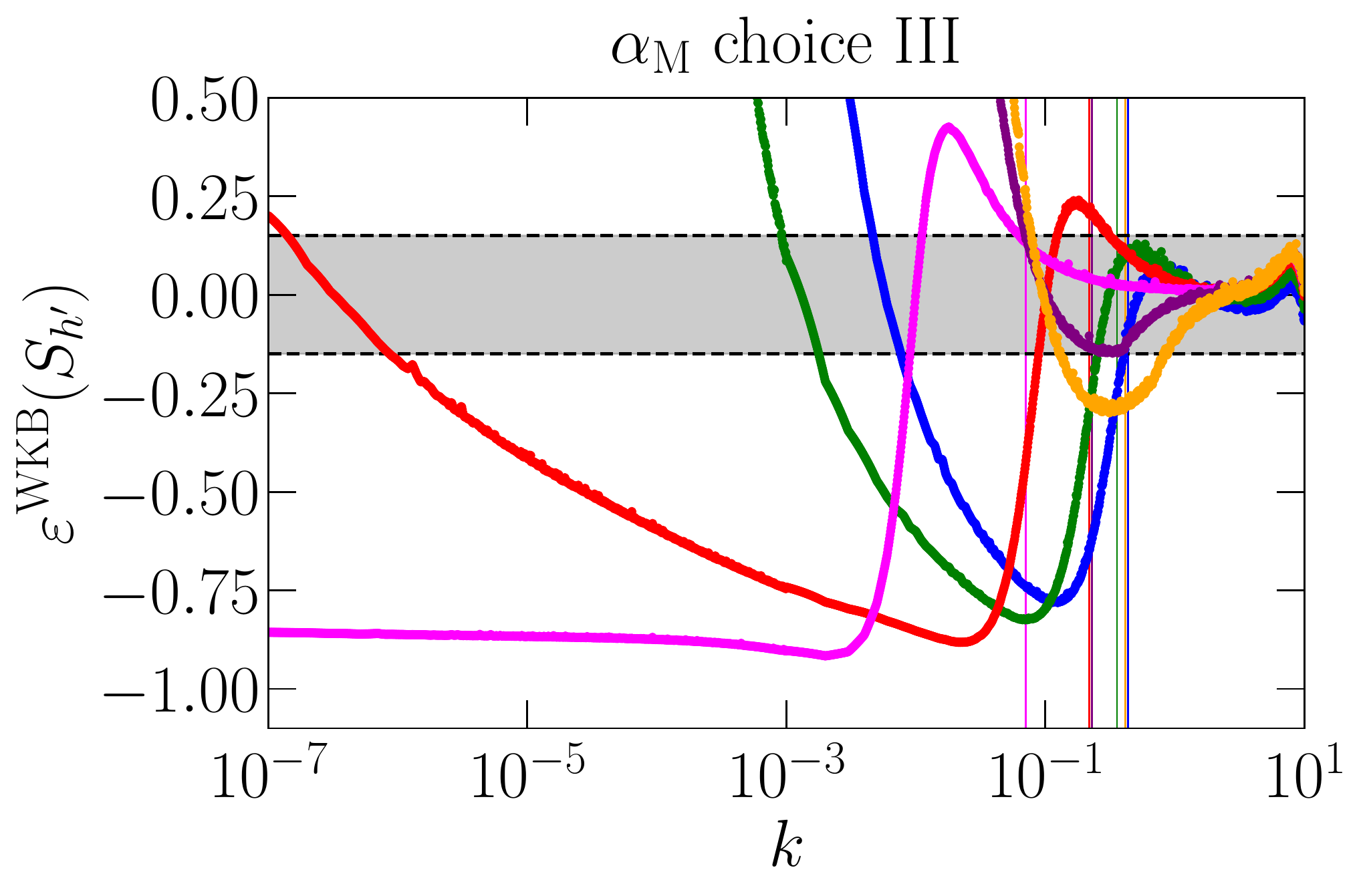}
\caption{
Relative error $\varepsilon^\WKB (S_{h'})$ in the saturated spectra at late times of the WKB approximation compared to the numerical simulations.
The vertical lines correspond to the wave number
$\klimaM$, below which the WKB estimate is expected to break down.
The gray shaded regions in both panels represent relative errors of $\leq15\%$.
Different values of $\alpMz$ and line colors are consistent with the ones used in \Figss{fig:ts_all}{fig:sp_all}.}
\label{fig:sp_error_WKB}
\end{figure}

Finally, to quantify the departures of the WKB approximation from the
numerical solutions, especially at $k\leq\klimaM$,
we show the relative errors in the saturated spectra as $\varepsilon^\WKB
(S_{h'})\equiv [S_{h'} (k)^\WKB - S_{h'} (k)]/S_{h'} (k)$ in
\Fig{fig:sp_error_WKB}.
We observe that for positive values of $\alpMz$, 
the errors grow monotonically as $k$ becomes smaller.
In fact, $\varepsilon^\WKB(S_{h'})\sim k^{-2}$ is found due to 
a consistent disagreement between the flat spectra predicted by the
WKB formalism and $k^{- \beta_0}$ with $\beta_0 \gtrsim 0$ obtained
numerically (see \Tab{tab:kcrit_beta}).
On the other hand, for negative values of $\alpMz$,
$\varepsilon^\WKB(S_{h'})$ also increase as $k$ becomes smaller.
But since $\alpMz < 0$ still produces a similar IR enhancement than that predicted by the WKB,
the slope in the error spectrum does not become as steep as $k^{-2}$ and depends 
on the corresponding values of $\beta_0$ in \Tab{tab:kcrit_beta}.
In all of the runs,
the errors are bounded by $\lesssim15\%$ at $k\gtrsim \klimaM$,
where the WKB estimate is expected to be valid.

\begin{table}[t]
    \centering
    \begin{tabular}{c|r|c|c|c}
        & $\alpMz$ & WKB & Choice 0 ($\alpM$) & Choice III ($\alpM$)  \\ \hline
        $k^\crit_{\beta=2}$ & $-0.5$ & $1.25 \times 10^{-1}$ & $1.32 \times 10^{-1}$ & $1.79 \times 10^{-1}$ \\
        & $-0.3$ & $7.5 \times 10^{-2}$ & $ 9.04 \times 10^{-2}$ & $1.20 \times 10^{-1}$ \\
        & $-0.1$ & $2.5 \times 10^{-2}$ & $3.83 \times 10^{-2}$ & $5.16 \times 10^{-2}$ \\
        & $-0.01$ & $2.5 \times 10^{-3}$ & $4.58 \times 10^{-3}$ & $6.63 \times 10^{-3}$ \\ \hline
        $k^{\crit,2}_{\beta=2}$ & $-0.5$  & 0 & $6.73 \times 10^{-2}$ & $9.93 \times 10^{-2}$ \\
        & $-0.3$  & 0 & $3.09 \times 10^{-2}$ & $5.54 \times 10^{-2}$ \\
        & $-0.1$  & 0 & $4.42 \times 10^{-3}$ & $9.62 \times 10^{-3}$ \\
        & $-0.01$ & 0 & $5.04 \times 10^{-5}$ & $7.44 \times 10^{-5}$ \\ \hline
      $\beta_0$ & $-0.5$  & $2$ & $1.38$ & $0.95$\\
      & $-0.3$  & $2$ & $1.66$ & $1.47$\\
      & $-0.1$  & $2$ & $1.89$ & $1.84$\\
      & $-0.01$ & $2$ & $2.00$ & $1.99$\\
      & $0.1$ & $2$ & $0.10$ & $0.15$\\
      & $0.3$ & $2$ & $0.30$ & $0.44$\\
    \end{tabular}
    \caption{Spectral features in the IR regime for the values
    of $\alpMz$ used in the numerical runs of \Tab{tab:run_params} and for choices 0 and III.
    $k^\crit_{\beta=2}$ and $k^{\crit, 2}_{\beta=2}$
    indicate the wave numbers at which the spectral slope becomes
    $-2$, which occurs at $k_{\beta=2} = \{0, \fourth |\alpM^*|\}$ for the WKB estimate.
    $\beta_0$ is the asymptotic slope in the numerical runs,
    such that $\xi (k) \sim k^{-\beta_0}$ at large scales,
    which corresponds to $\beta_0 = 2$ for the WKB estimate.}
    \label{tab:kcrit_beta}
\end{table}

\section{Observational implications}
\label{sec:obs_implications}

To infer the observational prospects of detecting a modified GW background, 
we convert the linear energy spectrum $S_{h'}(k)$ directly obtained from the code to the commonly used logarithmic energy spectrum $\OmGW(f)$ via \Eq{eqn:OmGW_Sh}.
As a result, the GR solution in the form of the double broken power law used in the numerical simulations,
i.e., $S_{h'}(k)\propto k^2$, $k^0$, and $k^{-{11\over3}}$ in the low, intermediate, and
high wave number regimes, given by the smoothed double broken power law in \Eq{sp_ini},
becomes $\OmGW(f)\propto f^3$, $f$, and $f^{-{8\over3}}$,
respectively.
The normalized wave numbers used in the code are converted to the
present-day physical frequencies via
\begin{equation}
f = \frac{\cT k\hhh_*}{2\pi} \, \frac{a_*}{a_0},
\label{f_k}
\end{equation}
where $\HH_*$ and $a_*/a_0$ are given in \Eqs{as_a0}{HHs} and depend
on the time at which the GWs are generated.
The term $\cT$ appears in \Eq{f_k} due to the modified dispersion relation at present time.
We have found that only $\alpM$ introduces spectral changes, so
we focus on the cases with $\cT = 1$.
However, modified theories of gravity with $\cT \neq 1$ would 
allow the resulting GW spectra $\OmGW (f)$ to be shifted in $f$ 
by $\cT$.

Assuming that the origin of the GWs is the EWPT and that the characteristic scale of the
source is $\HH_* \lambda_{\rm s} = 10^{-2}$ (given, for example,
by the mean separation of the broken-phase bubbles in a first-order EWPT \cite{Turner:1992tz}),
then $\OmGW(f)$ today peaks around the mHz band, which
corresponds to the peak sensitivity of the
Laser Interferometer Space Antenna (LISA).
On the other hand, if the GW signal is produced at the QCDPT with
a characteristic length scale of the order of the Hubble scale
$\HH_* \lambda_{\rm s} \approx 1$, then it would be compatible
\cite{RoperPol:2022iel}
with the recent observations by the different pulsar timing
array (PTA) collaborations \cite{NANOGrav:2020bcs, Goncharov:2021oub, Chen:2021rqp, Antoniadis:2022pcn} in the nHz
band of frequencies, which are, however, not yet confirmed to 
correspond to a GW background.
In particular, for a signal produced from MHD turbulence, the peak
is estimated to occur at $k_{\rm GW} \simeq 1.6 \times 2\pi/(\HH_* \lambda_{\rm s})$ \cite{RoperPol:2022iel}, which is obtained by
taking $k_* \simeq 0.875 \, k_{\rm GW}$ in \Eq{sp_ini}.

\begin{figure}[t]
\centering
\includegraphics[width=.73\textwidth]{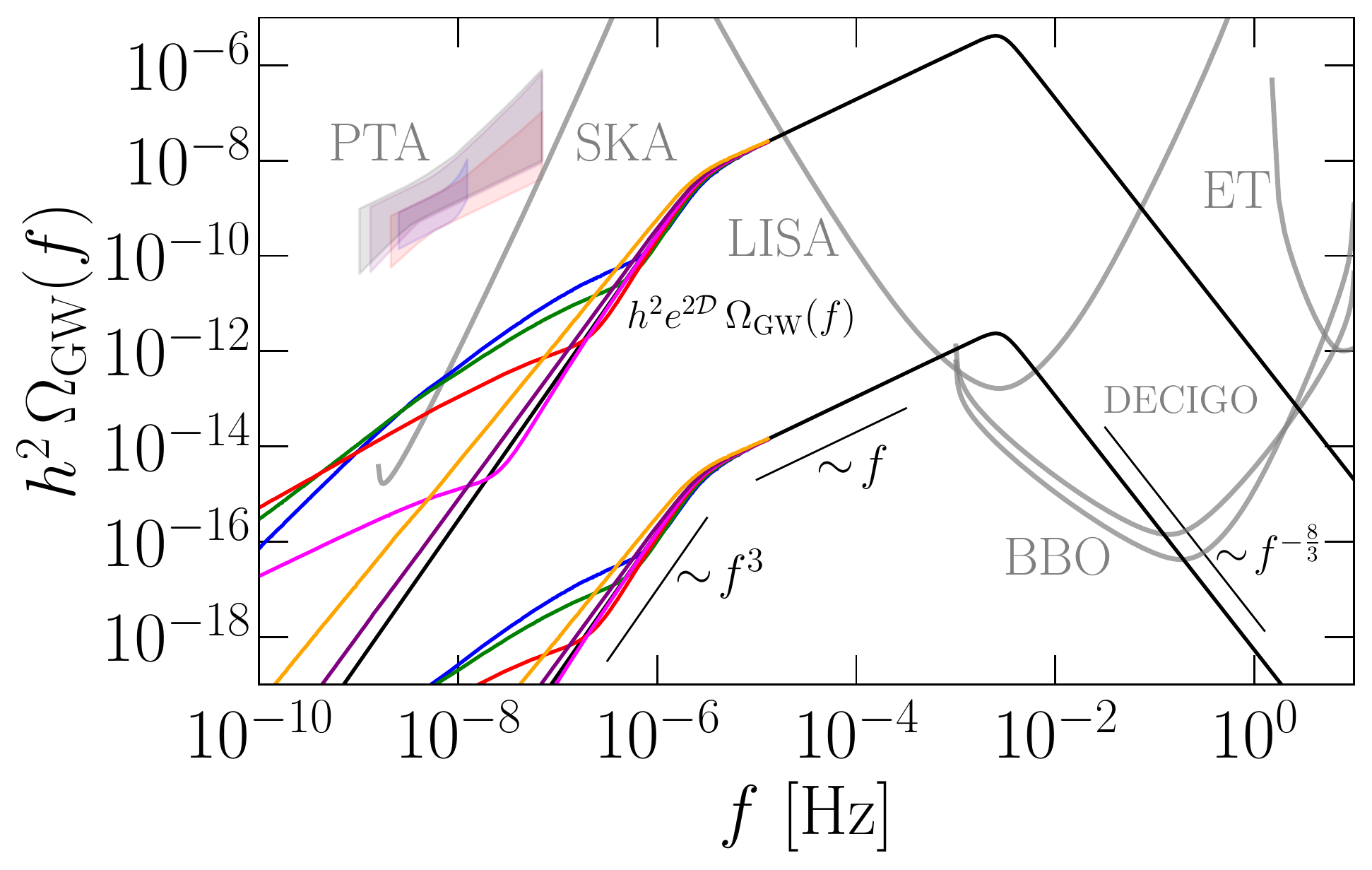}
\includegraphics[width=.75\textwidth]{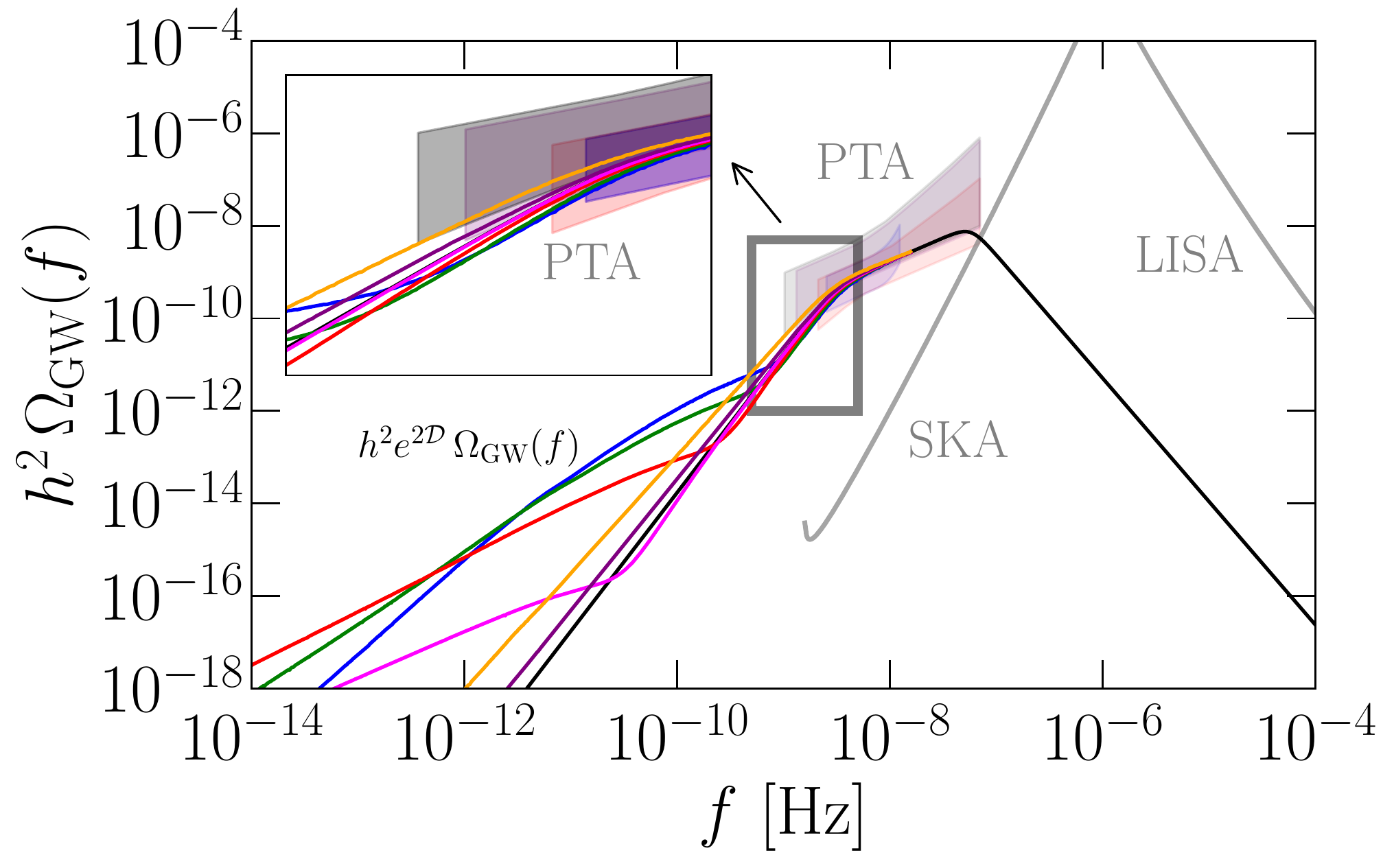}
\caption{Present-day modified GW energy spectrum for the numerical runs of series M3 (see \Tab{tab:run_params}) for
GW signals produced at the EWPT (upper panel) and at the
QCDPT (lower panel).
GR solution is shown in black.
Different values of $\alpMz$ and line colors are consistent with the ones used in \Figss{fig:ts_all}{fig:sp_all}.
Several detectors' PLIS curves for a signal-to-noise ratio of 10 are shown in gray for comparison.
The recent PTA reported observations are shown as shaded
regions in blue (NANOGrav), red (PPTA), purple (EPTA),
and black (IPTA) for their $2\sigma$ confidence amplitudes
in the range of slopes $\gamma \geq 0$ for
$\Omega (f) \sim f^\gamma$.}
\label{fig:OmGW_obs}
\end{figure}

Choosing the cases with the most pronounced modifications,
we show in \Fig{fig:OmGW_obs} the resulting $\OmGW (f)$ of the runs using
the choice III of $\alpM$ parameterizations, i.e., the runs of series M3; see \Tab{tab:run_params},
for the cases in which the GW signal is produced at the EWPT and
at the QCDPT.
For observational comparisons, 
we show the power law integrated sensitivity (PLIS) curves \cite{Schmitz:2020syl} of proposed 
future detectors such as Square Kilometer Array (SKA)~\cite{Moore:2014lga}, DECi-hertz Interferometer Gravitational wave Observatory (DECIGO) \cite{Seto:2001qf}, Big Bang Observer~\cite{Crowder:2005nr},
LISA~\cite{Caprini:2019pxz}, and Einstein Telescope (ET) \cite{Punturo:2010zz}.

\begin{figure}[t]
\centering
\includegraphics[width=.75\textwidth]{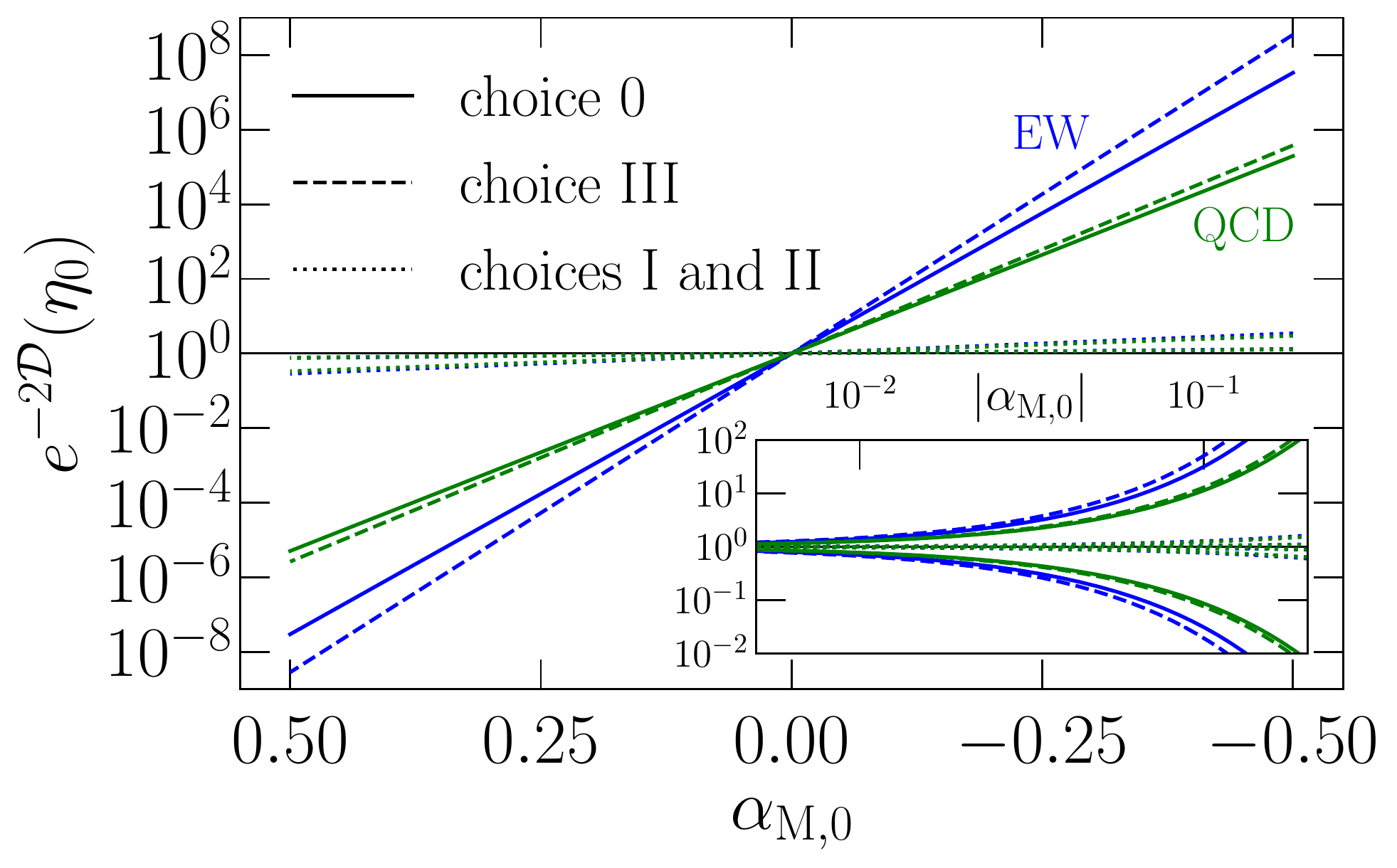}
\caption{Total amplification or depletion factor as a
function of $\alpMz$ for a GW signal produced at the EWPT (blue)
and at the QCDPT (green) for the different choices of $\alpM$ parameterization in \Eq{eqn:alpM_param}.}
\label{fig:OmGW_amplification}
\end{figure}

Due to the factor $e^{-2\ddd}$, negative values of $\alpMz$ boost 
the GW spectral amplitude, which increases the chance of GW detection.
Hence,
the (non-)observation of a GW signal could serve as a constraint 
on the value of $\alpMz$.
However, the present-day amplitude
of the GW spectrum is degenerate with respect to the value of $e^{-2\ddd}$
at present time (shown in \Fig{fig:OmGW_amplification}, it depends on 
$\alpMz$, the time-dependent parameterization of $\alpM$, and the time 
of generation of the GWs)
as well as the GW energy density produced at the time of generation $\EEGW^*$.
We show in \Fig{fig:OmGW_obs}, for different values of $\alpMz$,
the expected spectra at present time obtained
for fixed values $e^{-2\ddd} (\eta_0) \, \EEGW^* \simeq 2.5 \times 
10^{-7}$ and $4.44 \times 10^{-1}$ at the EWPT with 
$k_{\rm s} = 2 \pi/(\HH_*\lambda_{\rm s}) \simeq 
600$, which correspond to the spectral peaks at present time $h^2 \,
\OmGW \simeq 2.34 \times 10^{-12}$ and $4.15 \times 10^{-6}$ at
$2.5\,$mHz.
Note that the former amplitude would correspond to $q \eee_\turb^* 
\simeq 0.3$ in GR while the latter would be $q \eee_\turb^* 
\simeq 400$. This is completely unrealistic in GR
while in MG, it would again be compatible with $q \eee_\turb^* \simeq 0.3$ if $e^{-2\ddd} (\eta_0) \simeq 2 \times 10^6$ that
can be obtained if $\alpMz \approx -0.3$ 
(see choice III in \Fig{fig:OmGW_amplification}).
In this case, the GW signal could be amplified by MG such that the low-frequency tail could become detectable by
SKA and the different PTA collaborations in the future, in a way
that would break the degeneracy by a potential
observation over different frequency bands.
At the nHz frequencies, the $f^3$ branch
is modified to a different slope that depends on the value of
$\alpM^*$ at the time of generation, yielding a potential indirect observation
of the MG parameter during RD era.
The non-observation of such a signal at nHz frequencies
would similarly constrain the value of $\alpM^*$ if a compatible signal
is observed by LISA.
Otherwise, only the product $e^{-2\ddd} (\eta_0)\, \EEGW^*$ could be
constrained by not detecting such a signal in the LISA band.
Similarly, an even larger amplification $e^{-2\ddd} (\eta_0) \, \EEGW^*
\simeq 50$ could in principle allow the
large frequency branch $f^{-{8/3}}$ to fall into ET sensitivity.
Hence, ET could also be used to put upper bounds on the amplifications,
although less stringent than those combining LISA and SKA.

For the QCDPT, we take the fixed value $e^{-2\ddd} (\eta_0) \, \EEGW^* 
\simeq 5 \times 10^{-4}$ with $k_{\rm s} \simeq 10$, which corresponds to a
spectral peak $h^2 \, \OmGW \simeq 7.4 \times 10^{-9}$ at $50.6 \,$nHz.
In GR, this corresponds to $q \eee_\turb^* \simeq 0.22$ so it is compatible
with, for example, the primordial magnetic fields proposed in refs.~\cite{Neronov:2020qrl,Brandenburg:2021tmp,RoperPol:2022iel} in
connection to the reported PTA observations.
In this case, the horizon at the QCDPT scale is within the range of frequencies
where PTAs are sensitive and hence, the amplitude can be slightly modified
by the value of $\alpM^*$ at the time of generation.
Such modifications are small and at the moment, the data is not accurate
enough for such a precision measurement.
However, this is an interesting feature that occurs around the horizon and
the potential observation of a deviation with respect to the GR spectrum
around the horizon scale could be an indication of non-zero $\alpM^*$ around the
QCD scale.
On the other hand, the current PTA observations allow us to constrain
the product $e^{-2\ddd} \EEGW^*$ for different values of $k_{\rm s}$ since
we can find their values at which the GW signal would be larger than the reported
PTA common-noise observations: $e^{-2\ddd} \EEGW^* \leq 2.5 \times 10^{-3}$ for
$k_{\rm s} \approx 10$, $ \leq 2.5 \times 10^{-2}$ for
$k_{\rm s} \approx 60$, $\leq 4 \times 10^{-2}$ for $k_{\rm s} \approx 100$, 
$\leq 2.5 \times 10^{-1}$ for $k_{\rm s} \approx 600$.

In addition, the expected value of $\EEGW^*$ depends on the
mechanisms leading to GW production, in particular, their efficiency $q$, which
presents a fair amount of uncertainty.
Recall \Eq{eqn:EEGW_efficiency}, which states that the initial GW energy 
is determined by the initial turbulent source energy $\eee_\turb^*$, the 
production efficiency $q$, and the characteristic size of turbulent 
eddies $k_{\rm s}$.
Therefore, in practice, constraining $\alpMz$ via its amplification
effects can be rather challenging.

Finally, we note that large amplifications due to MG
could potentially
lead to GW signals produced at the EWPT that would dominate the sensitivity budget of various detectors, mostly LISA, but also
DECIGO and BBO,
for example if the product $e^{-2\ddd} (\eta_0) \, \EEGW^* \sim 10^{-1}$
(see \Fig{fig:OmGW_obs}).
Hence, at the moment that GW signals of astrophysical origin that
propagate over a shorter period of cosmic history are detected by LISA, 
DECIGO and/or BBO, one can put upper bounds to
$e^{-2\ddd} (\eta_0) \, \EEGW^*$.

\section{Conclusions}
\label{sec:conclusions}

GWs in modified gravity exhibit features in their
energy spectra different from what can be expected from GR.
We have explored the significance of such features in terms of the spectral
slopes and amplitudes in different frequency ranges, under different
time-dependent functional forms and values of two modification 
parameters --- the GW friction $\alpM$ and the tensor speed excess $\alpT$.

We have computed the expected GW energy density spectra using the WKB 
approximation, and have found that the GW energy density depends on the
time evolution of $\alpT$ and $\alpM$, while its spectral shape only
depends on $\alpM^*$ at the time of
GW generation,and the time dependence of $\alpT$.
However, a number of limiting scales below which the WKB approximation might
not hold have been
identified in \Eqss{kcritEW}{WKB_limitations},
motivating numerical studies to go beyond the WKB estimate,
especially at superhorizon scales.
For this purpose, we have performed numerical simulations using the
{\sc Pencil Code} to study the propagation of GWs along the
history of the universe from their time of generation, assumed to be
during the RD era, e.g., at the EWPT or at the QCDPT, until the present
time.

The current-day value of the tensor speed excess is already tightly constrained by the precise measurement of the speed of GWs to $\alpTz\lesssim\ooo(10^{-15})$.
In this paper, we have found that the modifications
on the GW energy density spectra introduced by 
a constant
$\alpT$ are negligible,
even when $\alpT$ takes larger values (as it might be allowed where
it is not constrained, e.g., in the past or at frequencies out of the 
LIGO--Virgo band).
In addition, $\alpT$ can modify the dispersion relation, leading
to a shift in $\OmGW (f)$ by $\cT$.
On the other hand, $\alpM$ effectively acts as an (anti-)damping term in the GW equation 
and can impart changes to the energy spectrum of GWs
with potential implications to future observations.
Considering sources of GWs at the EWPT and QCDPT,
the present-day GW spectrum is expected to peak in the mHz and nHz frequency bands, respectively.

Compared to the standard GW spectra obtained in GR,
the changes due to $\alpM$ are characterized by an overall boost or 
depletion factor $e^{-2\ddd}$, where $\ddd$ is given in \Eq{eqn:ddd_DelT}.
However, the GW energy density at present time due to modified gravity is degenerate with the produced density at the time of generation $\EEGW^*$.
Therefore, potential observations or non-observations of the GW
spectra by PTA, SKA, LISA, BBO, DECIGO, ET, or others, could only 
place bounds on the quantity $e^{-2\ddd}(\eta_0) \, \EEGW^*$ and not on $\alpM^*$ 
or $\EEGW^*$ separately if we only consider the amplitude at the peak of the GW spectrum.
For example, we have provided some upper bounds on $e^{-2\ddd}(\eta_0) \, \EEGW^*$ 
using the reported observations of a common-process spectrum over several
pulsars by the different PTA collaborations;  see \Sec{sec:obs_implications}.

In addition, the GW spectrum also presents
spectral modifications $\xi(k)$ around the horizon and at superhorizon scales.
If $\alpM^* < 0$, the modification leads to an enhancement of the 
GW spectrum $\xi(k)\sim k^{-\beta_0}$ with $\beta_0\in(0,2)$, which flattens the low-frequency part of the spectrum; see \Tab{tab:kcrit_beta}.
The enhancement found from the numerical simulations is different than
the one predicted by the WKB 
estimate ($\beta_0 = 2$) but it still
increases the detection prospects of the low-frequency
tail of the GW spectra.
We also find more moderate modifications to the spectral shape
around the horizon scale for all $\alpM^*$ and at all scales when $\alpM^* > 0$.
These modifications allow us to characterize the resulting spectra due to MG.

The degeneracy between $\alpM$ and 
$\EEGW^*$ can be broken by observing the 
signal at multiple frequency bands,
as $\alpM^*$ leads to a low-frequency spectral 
enhancement, while under GR
such enhancement does not appear.
This implies that, for example, a GW spectrum peaking in the mHz band can 
potentially have its tail in the nHz band, above PTA or SKA sensitivities.
The observation of such a low-frequency tail would then be a direct indication
of a non-zero $\alpM$ at the time at which the GW signal has been produced
since the slope and amplitude would only depend on $\alpM^*$.

Overall, a cosmological GW spectrum spanning many orders of magnitudes in frequency has the potential to constrain the parameter space of modified theories of gravity,
especially if the modifications occur in the early universe such as the RD era.

\paragraph{Data availability.} 
The source code used for the numerical solutions of this study,
the {\sc Pencil Code}, along with the additions included for the present study, is freely available~\cite{pencil};
see also ref.~\cite{DATA} for the numerical data.
The calculations and the routines generating the plots
are publicly available on GitHub\footnote{\url{https://github.com/AlbertoRoper/GW_turbulence/tree/master/horndeski}} \cite{GH}.

\acknowledgments
Support through the grant 2019-04234 from the Swedish Research Council
(Vetenskapsr{\aa}det) is gratefully acknowledged.
Nordita is sponsored by Nordforsk.
A.R.P. acknowledges support by the Swiss National Science Foundation (SNSF Ambizione grant 
 \href{https://data.snf.ch/grants/grant/208807}{182044}), the
French National Research
Agency (ANR) project MMUniverse (ANR-19-CE31-0020), and the Shota Rustaveli
National Science Foundation (SRNSF) of Georgia (grant FR/18-1462).
We acknowledge the allocation of computing resources provided by the
Swedish National Allocations Committee at the Center for Parallel
Computers at the Royal Institute of Technology in Stockholm,
and by the Grand Equipement National de Calcul Intensif (GENCI) to the project “Opening new windows
on Early Universe with multi-messenger astronomy”
(A0110412058).

\appendix

\section{Friedmann equations}
\label{sec:appendix_Friedmann}

Friedmann equations can be expressed in
terms of the energy density $\Om(a)$ and a smooth equation of state (EOS) $w(a)$ as
\begin{equation}
    \frac{\ddot a}{a} = - \frac{1}{2} H_0^2 \, \Om (a)
    \bigl[1 + 3 w (a)\bigr],
    \quad
    \biggl(\frac{\dot a}{a} \biggr)^2 =
    H_0^2 \, \Om(a),
    \label{friedmann2}
\end{equation}
where $\Om(a)$ is defined to be the ratio of total energy density to
the present-day critical energy density
$\rho_{\crit, 0}\equiv3H_0^2/\kap$, i.e.,
\begin{align}
\Om(a) = \frac{\rho (a)}{\rho_{\crit, 0}} = &\,
    \Omega_\rad (a) + \Omega_\mat (a) + \Omega_{\Lam,0} \nonumber \\
    = &\,
     \biggl({a\over a_0}\biggr)^{-4} \,
    \frac{g_*}{g_*^0} \biggl(\frac{g_{\rm S}}{g_{\rm S}^0}
    \biggr)^{-{4\over3}} \, \Omega_{\rad,0} + \biggl({a\over a_0}\biggr)^{-3} \, \Omega_{\mat,0} + \Omega_{\Lambda,0}.
    \label{rho_a}
\end{align}
Numerically, we take the present-time values of $\Om_{\Lam,0} \simeq 0.684$,
$\Om_{\rad,0} \simeq 4.16 \times10^{-5} \, h^{-2}$, and
$\Om_{\mat,0} = 1 - \Om_{\rad, 0} - \Om_{\Lam, 0}
\simeq 0.316$, where $h$ takes into account the uncertainties on
the present-time Hubble rate $H_0 = 100 \, h$ km/s/Mpc.
We set $h \simeq 0.67$ for the numerical
studies, using the value observed by CMB experiments \cite{Planck:2018vyg},
and $g_*^0 \simeq 3.36$ and $g_{\rm S}^0 \simeq 3.91$ are 
the reference
relativistic and adiabatic degrees of freedom at the present time.\footnote{Note that
neutrinos' contribution to the radiation energy density is 
accounted for by taking
\begin{equation}
    g_*^0 = 2 \Biggl[1 + N_{\rm eff} \frac{7}{8} \biggl(
    \frac{4}{11}\biggr)^{4\over3}\Biggr] \simeq 3.363, \nonumber
\end{equation}
where $N_{\rm eff} \approx 3$,
instead of $g_*^0 = 2$ at the present day due to photons only.
This leads to an excess in the calculation of the
radiation energy density after neutrinos become massive.
However, this occurs when the radiation energy is subdominant and
hence, it does not affect our calculations.}

The evolution of the relativistic and adiabatic degrees of freedom
as a function of the temperature during RD are taken
from ref.~\cite{Hindmarsh:2020hop} and expressed as a function
of $a$ by taking $a^3 \, T^3 \gS$ to be constant, following an
adiabatic expansion of the universe.

The characteristic EOS, corresponding to
the energy density of \Eq{rho_a}, can be computed combining
equations~(\ref{friedmann2})
by taking the time derivative of the second equation and
introducing the first equation.
This yields
\begin{gather}
        \frac{\dot \Om (a)}{\Om (a)} = 
    -3 (1 + w) \frac{\dot a}{a},
    \label{Om_a_dot}
\end{gather}
which allows us to compute $w (a)$ using \Eq{rho_a},
\begin{equation}
    w (a)= \bigg({1 \over 3} \Om_{\rad} (a) 
    - \Om_{\Lam, 0}\bigg) \, \Om^{-1} (a).
    \label{eqn:w_a}
\end{equation}
\EEq{Om_a_dot} justifies
the expected evolution of $\Om(a)$ used in \Eq{rho_a}: approximately 
proportional to $a^{-4}$ during 
the RD era, to $a^{-3}$ during
the MD era, and
constant during $\Lambda$D.
During RD, the evolution of the degrees of freedom induces some
modifications with respect to the $a^{-4}$ evolution.
\EEq{eqn:w_a} yields
$w = 1/3$ for RD, $w = 0$ for MD, and
$w = -1$ for $\Lambda$D, as expected.
In the intermediate times, the functions $\Om(a)$ and $w (a)$ transition
smoothly.
We can express \Eq{friedmann2} in terms of the
normalized conformal time $\eta \HH_*$ as
\begin{equation}
   \HH = \frac{H_0}{\HH_*} \, a \sqrt{\Omega (a)},
    \quad 
    \frac{a''}{a} = \frac{1}{2} \HH^2
    \bigl[1 - 3 w (a) \bigr],
    \label{eqn:a_derivatives_omega_appendix}
\end{equation}
which corresponds to \Eq{eqn:a_derivatives_omega}.
Finally, since \Eq{eqn:GW_mod_chi} is expressed in terms of conformal time $\eta$ but \Eq{eqn:a_derivatives_omega_appendix} is still written in terms of $a$, we 
would like to substitute the variables via $a(t)$ or $a(\eta)$, which can be obtained using \Eq{friedmann2}:
\begin{equation}
    \dot{a} = H_0 \, a \sqrt{\Omega (a)} \Rightarrow
    \dd \bigl(H_0 t\bigr) = \frac{\dd a}{a \sqrt{\Omega (a)}}
    \Rightarrow H_0 \bigl( t - t_{\rm ini} \bigr) =
    \int_{a_{\rm ini}}^a
    \frac{\dd a}{a \sqrt{\Omega (a)}}.
    \label{dota}
\end{equation}
This allows us to compute $t (a)$ and then invert the
relation to obtain $a (t)$.
Similarly, in conformal time, we have $\dd t = a \dd \eta$, so we solve
\begin{equation}
    H_0 \bigl( \eta - \eta_{\rm ini} \bigr) =
    \int_{a_{\rm ini}}^a
    \frac{\dd a}{a^2 \sqrt{\Omega (a)}}.
    \label{dota}
\end{equation}
For the numerical integration, we set $a_{\rm ini} = 10^{-20}$ at
$t_{\rm ini} = \eta_{\rm ini} = 0$, which yields accurate results
for all $a \gtrsim 10^{-19}$.
Further details and numerical results to Friedmann equations can
be found on the \texttt{GW\_turbulence} \href{https://github.com/AlbertoRoper/GW_turbulence/tree/master/cosmology}{GitHub} project \cite{GH}.

In the present work, we consider that the GWs are generated
during a phase transition
(in particular, at the EWPT or QCDPT) within the RD era.
Assuming adiabatic expansion of the universe, one can compute $a_*/a_0$ as a function of the temperature $T_*$ and the adiabatic degrees of freedom,
\begin{equation}
    \frac{a_*}{a_0} = \frac{T_0}{T_*} \biggl(\frac{\gS^0}
    {\gS} \biggr)^{1\over3} \simeq 7.97 \times 10^{-16}
    \frac{T_*}{100 \, \GeV} \biggl(\frac{\gS}{100}\biggr)^{-{1\over3}},
    \label{as_a0}
\end{equation}
where we take $T_0 \simeq 2.7255$\,K.
Setting $a_* = 1$, such that $\HH_* = H_*$,
one gets the following value of the Hubble rate (valid during the RD era),
\begin{equation}
    \HH_* = \frac{\pi T_*^2}{3} \sqrt{\frac{\kappa \,g_*}{10}} \simeq 2.066 \times 10^{10} \Hz \, \biggl(\frac{T_*}{100 \GeV}
    \biggr)^2 \biggl(\frac{g_*}{100} \biggr)^{1\over2},
    \label{HHs}
\end{equation}
with\footnote{During RD with $w = {1\over3}$, $\eta = 1/\HH$ holds exactly, but the dependence of the
radiation energy density with $g_*$ and $\gS$ induce small deviations.}
$\eta_* \approx \HH_*^{-1}$.
These results allow us to use the solutions from Friedmann equations
and adapt them to compute the variables that appear in \Eqs{eqn:GW_mod}{eqn:GW_mod_chi}, normalized to the specific epoch of GW generation.

\section{GW energy spectrum via WKB approximation}
\label{sec:WKB_spec}

In this section, we present the derivation of the GW spectrum obtained using the WKB approximation introduced in \Sec{GW_sp_WKB}.
We start with the WKB solution for $\chi = e^{\ddd} h$, given in \Eq{eq_chi_modGW_hij}, and assume that at the initial time, the solution is in the free-propagating regime and follows GR\footnote{We neglect here the effects of MG during the GW production and focus on the MG effects after the GWs start to propagate. Otherwise, $|h'_*(k)| \neq k |h_*(k)|$ in general and the final GW spectrum would present some additional terms.
This scenario is left for future work.}, such that $|h'_*(k)| = k |h_*(k)|$,
\begin{equation}
\chi(k, \eta) = e^{-\tilde\ddd} h_*' (k) \Bigg[\frac{1}{k}
\cos k\theta + 
\frac{1}{k \cT^*} \biggl(1 + \frac{{\cal A}_*}{k} \biggr)
\sin k\theta\Biggr],
\end{equation}
where $\theta = \tilde c_{\rm T} (\eta\hhh_* - 1)$ and ${\cal A}_* 
=\half(\alpM^* + {{\cT'}^* \over \cT^*})$.
Its time derivative is
\begin{align}
    \chi' (k, \eta) = &\,
    e^{-\tilde \ddd} h_*' (k)
    \biggl[\frac{\cT}{\cT^*}\biggl(1 + \frac{{\cal A}_*}{k}
    \biggr) \cos k \theta - 
    \cT \sin k \theta \biggr]  - \frac{\cT'}{2 \cT} \chi(k, \eta).
    \label{chi_der}
\end{align}
We have used that, for any function $f$, $[f (k\theta)]' = k \theta' f'(k \theta) = \cT \, k \, f(k \theta)$, since $\theta' = \tilde c_{\rm T} + \tilde c_{\rm T}' (\eta \HH_* - 1) = \cT$, and $\tilde \ddd' = \cT'/(2 \cT)$.
The spectrum of GWs in normalized variables, given in \Eq{eqn:OmGW_Sh}, is proportional to
\begin{equation}
     \bra{h'^2 (k, \eta)} = e^{-2\ddd (\eta)}
     \Bra{\bigl[\chi' (k, \eta) - \ddd' (\eta) \chi (k, \eta)\bigr]^2},
\end{equation}
where the brackets indicate an average over oscillations in time.
Taking into account that $\ddd' = \half \alpM \HH$
and using \Eq{chi_der}, the term in brackets is
\begin{align}
    \chi' (k, \eta) - {\cal D}' \chi (k, \eta) =
    e^{-\tilde \ddd} h_*' (k) \Bigl[{\cal C}_1 (k)
    \cos k \theta - {\cal C}_2 (k) \sin k\theta \Bigr],
    \label{chi_prime_1}
\end{align}
where we have defined
\begin{equation}
    {\cal C}_1 (k) = \frac{\cT}{\cT^*} \biggl(1 + \frac{{\cal A}_*}{k} \biggr) - \frac{1}{2k} \biggl(
    \frac{\cT'}{\cT} + \alpM \HH\biggr), \quad
    {\cal C}_2 (k) = \cT + \frac{1}{2 k \cT^*}
    \biggl(\frac{\cT'}{\cT} + \alpM \HH\biggr)
    \biggl(1 + \frac{{\cal A}_*}{k} \biggr).
\end{equation}
Hence, taking the square of \Eq{chi_prime_1} and averaging over oscillations in time we find
\begin{equation}
    \Bra{\bigl[\chi'(k, \eta) - {\cal D}' \chi (k, \eta)
    \bigr]^2} = \half e^{-2 \tilde \ddd} \bra{h_*' (k) \, h_*' (k)} \Bigl[{\cal C}_1^2 (k) + {\cal C}_2^2 (k) \Bigr],
\end{equation}
where
\begin{align}
    {\cal C}_1^2 (k) + {\cal C}_2^2 (k) =
    &\,
    \cT^2 \biggl(1 + \frac{1}{{\cT^*}^2} \biggr) + \frac{1}{k^2} \biggl[
    \frac{\cT^2 {\cal A}_*^2}{{\cT^*}^2} + \frac{1}{4}
    \biggl(1 + \frac{1}{{\cT^*}^2}\biggr)\biggl(
    \frac{\cT'}{\cT} + \alpM \HH \biggr)^2\biggr] \nonumber \\
    &\, +
    \frac{2 \cT^2 {\cal A}_*}{k {\cT}^{*2}} + \frac{{\cal A}_*}{2 k^3 {\cT^*}^2} \biggl(1 +
    \frac{{\cal A}_*}{2 k} \biggr)
    \biggl(\frac{\cT'}{\cT} + \alpM \HH \biggr)^2.
\end{align}
This expression allows us to compute the GW spectrum under the
MG theory from the time evolution of 
$\alpM$ and $\cT$, and, in particular, from these values at the time of GW production,
\begin{equation}
    \OmGW (k, \eta) = e^{-2 (\ddd + \tilde \ddd)} \,
    \OmGW^{\rm GR} (k) \, \xi(k, \eta),
\end{equation}
where $\xi(k, \eta)$ characterizes the GW spectrum in MG
with respect to the one expected under GR,
\begin{align}
    \xi(k, \eta)
    = &\,
    \frac{\cT^2}{2} \biggl(1 + \frac{1}{{\cT^*}^2}
    \biggr) + \frac{\cT^2}{2 k {\cT^*}^2}\biggl(
    \alpM^* + \frac{{\cT'}^*}{\cT^*}\biggr) + \frac{1}{8 k^3 {\cT^*}^2} \biggl(\alpM^* + 
    \frac{{\cT'}^*}{\cT^*} \biggr)\biggl(\frac{\cT'}{\cT}
    + \alpM \HH\biggr)^2
    \nonumber \\
    + &\, \frac{1}{8 k^2} \Biggl[ \frac{\cT^2}{{\cT^*}^2}
    \biggl(
    \alpM^* + \frac{{\cT'}^*}{\cT^*}\biggr)^2 +
    \biggl(1 + \frac{1}{{\cT^*}^2}\biggr)\biggl(
    \frac{\cT'}{\cT} + \alpM \HH\biggr)^2 \Biggr]
    \nonumber \\ + &\,
    \frac{1}{32 k^4 {\cT^*}^2}  \biggl(\alpM^* + 
    \frac{{\cT'}^*}{\cT^*} \biggr)^2 \biggl(\frac{\cT'}{\cT}
    + \alpM \HH\biggr)^2.
    \label{general_xi_modGR}
\end{align}
If we set $\cT^* = \cT$ and $\cT' = {\cT'}^* = 0$, we recover \Eq{eqn:EGW_mod}
\begin{equation}
    \xi (k, \eta) = 1 + \frac{1}{2}\alpT
    + \frac{\alpM^*}{2 k}
    + \frac{1}{8 k^2} \Biggl[
    \alpM^2\HH^2\biggl(1 + \frac{1}{\cT^2}\biggr) 
    + \alpM^{*2}
    \Biggr]
    + \frac{\alpM^*\alpM^2\HH^2}{8 k^3\cT^2}
    + \frac{\alpM^{*2}\alpM^2 \HH^2}{32 k^4 \cT^2}.
\end{equation}

\section{Numerical scheme}
\label{sec:numerical_scheme}

Following refs.~\cite{RoperPol:2018sap,Brandenburg:2021pdv},
we implement a matrix solver for the modified GW equation.
\EEq{eqn:GW_mod} can be expressed as
\begin{equation}
h_\ij'' + \sig h_\ij' + \om^2h_\ij = 0,
\label{eqn:GW_mod_rewrite}
\end{equation}
where
\begin{equation}
\sig\equiv\alpM\hhh\,,\,\,
\om^2\equiv\cT^2k^2 - \alpM\hhh^2 - \frac{a''}{a}.
\end{equation}
Then, assuming a solution of the type $\hij = e_\ij A e^{\lambda\eta}$ gives
the characteristic equation
\begin{equation}
\lam^2 + \sig\lam + \om^2 = 0.
\end{equation}
The eigenvalues $\lam$ can be obtained as
\begin{equation}
\lam_{1,2} = -\frac{1}{2}\Big(\sig\mp\sqrt{\sig^2 - 4\om^2}\Big).
\end{equation}
For $\del\eta\ll1$, 
we approximate the solution by assuming that $\lambda_{1,2}' \approx \lambda_{1,2}'' \approx 0$ during the time interval within time steps of the numerical solver.
Then, the solution for the strains in \Eq{eqn:GW_mod_rewrite} takes the form
\begin{align}
h_\ij(\eta + \del\eta) &
= \ccc_\ij e^{\lam_1\del\eta} + \ddd_\ij e^{\lam_2\del\eta}, \\
h_\ij'(\eta + \del\eta) &
= \ccc_\ij\lam_1 e^{\lam_1\del\eta} + \ddd_\ij\lam_2 e^{\lam_2\del\eta},
\end{align}
where $\ccc_\ij$ and $\ddd_\ij$ are constant amplitude coefficients evaluated at $\eta$.
Equivalently, in matrix form,
the solution can be rewritten as
\begin{equation}
\begin{pmatrix}
h \\
h'
\end{pmatrix}_\ij^{\eta + \del\eta}
=
\begin{pmatrix}
e^{\lam_1\del\eta} & e^{\lam_2\del\eta} \\
\lam_1e^{\lam_1\del\eta} & \lam_2e^{\lam_2\del\eta}
\end{pmatrix}
\begin{pmatrix}
\ccc \\
\ddd
\end{pmatrix}_\ij.
\end{equation}
Hence, the amplitude coefficients can be obtained as
\begin{align}
\begin{pmatrix}
\ccc \\
\ddd
\end{pmatrix}_\ij
& =
\begin{pmatrix}
e^{\lam_1\del\eta} & e^{\lam_2\del\eta} \\
\lam_1e^{\lam_1\del\eta} & \lam_2e^{\lam_2\del\eta}
\end{pmatrix}^{-1}
\begin{pmatrix}
h \\
h'
\end{pmatrix}_\ij^{\eta + \del\eta}\\
& = 
\frac{1}{(\lam_2 - \lam_1)e^{(\lam_1 + \lam_2)\del\eta}}
\begin{pmatrix}
\lam_2e^{\lam_2\del\eta} & -e^{\lam_2\del\eta} \\
-\lam_1e^{\lam_1\del\eta} & e^{\lam_1\del\eta}
\end{pmatrix}
\begin{pmatrix}
h \\
h'
\end{pmatrix}_\ij^{\eta + \del\eta},
\end{align}
and, in the limit $\delta \eta \rightarrow 0$, we get
\begin{align}
\lim_{\del\eta\map0}
\begin{pmatrix}
\ccc \\
\ddd
\end{pmatrix}_\ij
& =
\frac{1}{\lam_2 - \lam_1}
\begin{pmatrix}
\lam_2 & -1 \\
-\lam_1 & 1
\end{pmatrix}
\begin{pmatrix}
h \\
h'
\end{pmatrix}_\ij^{\eta},
\end{align}
where we assume that $\ccc_\ij$ and $\ddd_\ij$ are time-independent within
time steps.
Therefore, the time evolution of the relevant quantities 
can be obtained via a matrix multiplication as
\begin{equation}
\begin{pmatrix}
h \\
h'
\end{pmatrix}_\ij^{\eta + \del\eta}
=
\mmm
\begin{pmatrix}
h \\
h'
\end{pmatrix}_\ij^{\eta},
\label{eqn:GW_evol_mtx_GAFD}
\end{equation}
where
\begin{equation}
\mmm =
\frac{1}{\lam_1 - \lam_2}
\begin{pmatrix}
\lam_1e^{\lam_2\del\eta} - \lam_2e^{\lam_1\del\eta} 
& e^{\lam_1\del\eta} - e^{\lam_2\del\eta} \\
\lam_1\lam_2(e^{\lam_2\del\eta} - e^{\lam_1\del\eta})
& \lam_1e^{\lam_1\del\eta} - \lam_2e^{\lam_2\del\eta}
\end{pmatrix}.
\label{eqn:evol_mtx_full}
\end{equation}
In the case of $\sig = 0$ and $\om^2 = k^2$,
the eigenvalues become $\lam_{1,2} = \pm ik$ 
and the evolution matrix reduces to the one relevant for GR,
\begin{equation}
\mmm = 
\begin{pmatrix}
\cos{k\del\eta} & \sin{k\del\eta}\\
-\sin{k\del\eta} & \cos{k\del\eta}
\end{pmatrix}.
\label{eqn:evol_mtx_simp}
\end{equation}

\section{Numerical accuracy}
\label{sec:numerical_accuracy}

\begin{figure}[t]
\centering
\includegraphics[width=\textwidth]{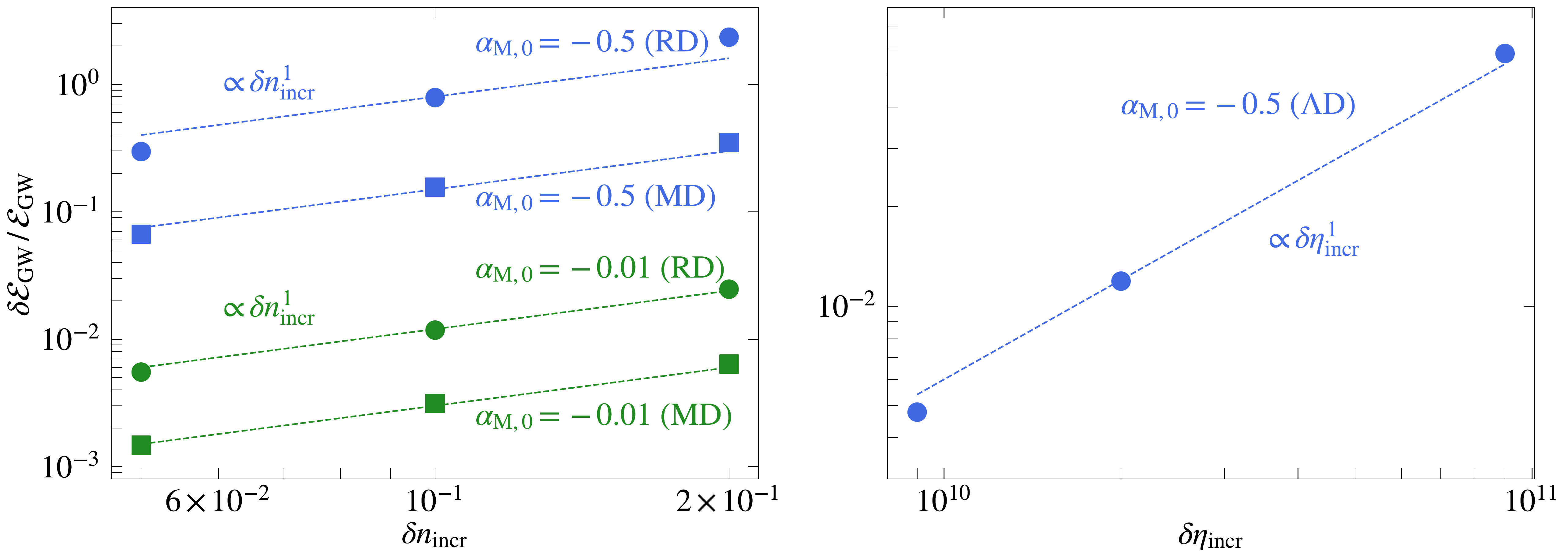}
\caption{Relative errors as a function of the numerical time step.
\textit{Left panel}: Accuracy during RD (round dots) and MD (square
dots) for $\alpMz = -0.5$ (blue) and $\alpMz = -0.01$ (green), where
$\del n_\incr$ indicates the difference in the time steps such
that $\delta \eta = \eta\, \delta n_\incr$.
\textit{Right panel}: Accuracy during $\Lam$D with $\del \eta = \del\eta_\incr$ directly indicating the linear time interval.
The errors are calculated relative to the values obtained from 
using the smallest $\delta n_\incr = 10^{-2}$ (left panel)
and $\del\eta_\incr = 5\times10^9$ (right panel).}
\label{fig:dEEGW_dn_incr}
\end{figure}

In this section, we study the numerical accuracy of the runs.
The GW solver in the {\sc Pencil Code} has been made accurate to second order 
in time in the presence of a source $T_\ij$ under GR.
However, the GW equation under modified theories of gravity lead to an
equation that can be approximated within small but finite time steps under
the assumption that the coefficients are constant in time,
leading to the numerical solver described in \Sec{sec:numerical_scheme}.
This method is accurate up to first order in the duration of the time step.
As the coefficient variations in time become smaller, the error is
also smaller, but the accuracy of the solver is still of first order.
This is demonstrated in \Fig{fig:dEEGW_dn_incr},
where we show on the left panel the relative errors in the GW energy during RD and MD,
and on the right panel the error during $\Lam$D.
Recall that, as introduced in \Sec{init_conds}, during RD and MD, we use increasing time steps, represented by $n_\incr$ here.
During $\Lam$D, we use linear time steps, with the interval
shown as $\eta_\incr$ in \Fig{fig:dEEGW_dn_incr}.
\begin{figure}[t]
\centering
\includegraphics[width=0.65\textwidth]{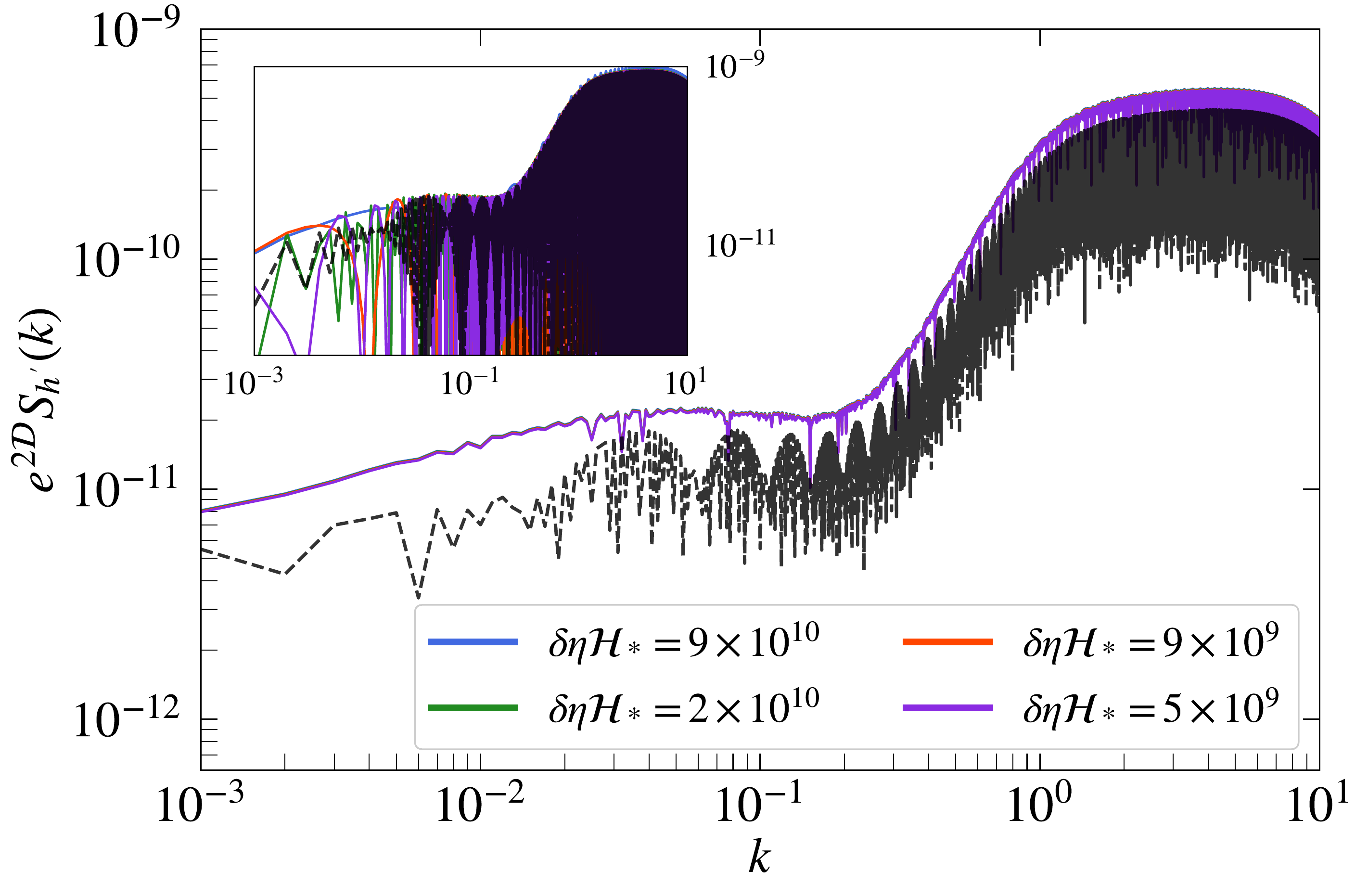}
\caption{Saturated energy spectrum $e^{2\ddd}S_{h'}(k)$ for $\alpMz = -0.5$ (run M0A) with different time steps $\del\eta\,\hhh_*$ during $\Lam$D,
where $n_\incr = 0.01$ is used during RD and MD.
The black dashed curves correspond to the run M0A performed entirely, i.e., including
$\Lam$D, with increasing time steps with $n_\incr = 0.01$.
The inset directly shows the spectrum at the final time without 
averaging over time oscillations.}
\label{fig:EGW_deta}
\end{figure}

We also test the accuracy of the spectra in \Fig{fig:EGW_deta},
where we show the saturated final energy spectrum for run M0A.
Different colors indicate the linear time intervals during $\Lam$D and
the black curve is the spectrum obtained by running M0A
entirely with increasing time steps using $n_\incr = 0.01$.
In other words, the colored curves only differ from the black one 
during $\Lam$D.
We observe that all of the chosen values of $\del\eta$ produce converging
energy spectra after averaging over oscillations in time,
although larger values of $\del\eta$ result in more fluctuations
in the numerical spectra (see the inset in \Fig{fig:EGW_deta}).
We also note that, compared to results from hybrid time steps, the
entirely nonuniform time steps somewhat underestimate the final energy spectrum.

These accuracy tests justify our choice to use nonuniform time steps during RD and MD to improve
the efficiency of the simulations, and uniform linear time steps (with
$\delta \eta$ that already gives converged results) during $\Lambda$D.

\bibliographystyle{JCAP}
\bibliography{references}

\end{document}